\documentclass[
  aps,
  prl,
  reprint,
  superscriptaddress,
  nofootinbib,
  longbibliography,
  floatfix]{revtex4-2}

\usepackage[T1]{fontenc}
\usepackage{lmodern}
\usepackage{microtype}
\usepackage{amsmath,amssymb,mathtools}
\usepackage{bm}
\usepackage{graphicx}
\usepackage{xcolor}
\usepackage[normalem]{ulem}
\usepackage{hyperref}

\hypersetup{
  colorlinks=true,
  linkcolor=blue!45!black,
  citecolor=blue!45!black,
  urlcolor=blue!45!black,
  pdftitle={Heavy fermion and superconductivity near Mott transition in twisted bilayer graphene}
}

\definecolor{ccgreen}{rgb}{0.00,0.45,0.20}

\definecolor{codexpurple}{RGB}{145,35,165}

\newcommand{\ii}{\mathrm{i}}
\newcommand{\dd}{\mathrm{d}}

\newcommand{\identity}{I}
\newcommand{\kstar}{k_*}
\newcommand{\Tr}{\operatorname{Tr}}
\providecommand{\ket}[1]{\lvert#1\rangle}

\begin{document}

\title{Emergent heavy fermion and superconductivity near Mott transition in twisted bilayer graphene}

\author{Ya-Hui Zhang}
\affiliation{Department of Physics and Astronomy, Johns Hopkins
University, Baltimore, Maryland 21218, USA}

\begin{abstract}
Near a bandwidth-tuned Mott transition, the Fermi velocity $v_F$ and quasiparticle residue $Z$ of a metal often vanish. Here, we show that analogous phenomena emerge in twisted bilayer graphene (TBG) at integer fillings and can be captured by an emergent heavy-fermion framework within a projective active-band limit. Unlike models incorporating remote bands, our effective heavy-fermion description arises from \textit{mixed-valence Mott} physics via decoupled charge and local-moment sectors. In the charge sector, active bands $c(\mathbf{k})$ hybridize with an emergent \emph{orthogonal fermion} $\psi(\mathbf{k})$ to open a large Mott gap at $|\mathbf{k}| > k_*$ ($k_*$ sets the momentum-patch size) and a quadratic band-touching semimetal near $\mathbf{k}=0$ at neutrality ($\nu=0$). The orthogonal fermion is a linear combination of the doublon and holon excitations and may be written as $\psi_i \sim (\delta n^f_i+\frac{1}{2})^{-1} f_i$. An emergent Kondo coupling $J_K \sim U$ ($U$ is the local Hubbard interaction) between $\psi$ and local moments $\psi'$ frames the Mott transition as a Kondo screening transition, tuned by the twist angle $\theta$. Away from the magic angle, a Kondo-screened heavy semimetal develops below $T_K$ (the Kondo temperature) with vanishing  $Z$. Introducing anti-Hund's coupling $J_A$ generates an s-wave fully gapped or nematic, nodally gapped superconducting dome near the Mott boundary even at $\nu=0$. At other integer fillings $\nu = \pm 1, \pm 2$, increasing bandwidth first drives the small-gap Mott state into an intermediate quadratic band-touching semimetal before entering a heavy Fermi liquid with large Fermi surfaces.  Our results establish a unified framework for emergent heavy fermion physics with both itinerant carriers and local moments from the $f$ orbital.
\end{abstract}

\maketitle

\emph{Introduction.---}
How a metal dies at a bandwidth-tuned Mott transition is a
central question of correlated matter.  In the
canonical Brinkman--Rice and dynamical-mean-field picture
\cite{BrinkmanRice1970,GeorgesRMP1996,ImadaRMP1998,FlorensGeorges2004,Senthil2008},
the quasiparticle residue $Z$, renormalized Fermi velocity
$v_F$, and coherence scale collapse as the interaction-to-bandwidth
ratio grows.  The difficulty is to capture the itinerant and localized nature of the phase simultaneously within a single framework.  Heavy-fermion materials
\cite{Doniach1977,SiSteglich2010,Lohneysen2007}
avoid this problem because localized moments
and the conduction sea occupy different orbitals.  For a one-orbital model, a Mott insulator with full localization can happen at integer filling, and 
 continuous Mott transitions have been observed in semiconductor-based moir\'e materials
\cite{LiMott2021,Ghiotto2021}. Then one may focus on the local moments and ignore the charge degree of freedom entirely.   On the other hand, in a topological band, full Mott localization is obstructed and we expect the presence of both the itinerant electrons and local moments even at integer filling. 

Magic-angle twisted bilayer
graphene (TBG) sharpens this question: its flat bands
\cite{BistritzerMacDonald2011} host
correlated insulators at integer fillings $\nu$ and adjacent
superconducting domes
\cite{CaoInsulator2018,CaoSC2018,Yankowitz2019,Lu2019,Oh2021,CaoNematic2021},
but a topological (Wannier) obstruction prevents a conventional
Hubbard description of the active bands
\cite{Po2018,Song2019}.  The topological heavy-fermion model (THFM)
resolves it by decomposing the active bands into local $f$
orbitals hybridized with topological conduction electrons $c$
\cite{SongBernevig2022,ShiDai2022} and motivates Kondo-lattice treatments 
\cite{ChouDasSarma2023,HuTsvelik2023,LauColeman2023,ZhouSong2024}
and dynamical mean-field studies \cite{Datta2023,Rai2024,Youn2024}.
However, their itinerant electrons lie mainly in the remote
bands. In the limit where the Hubbard $U$ is smaller than the band gap, the physics is closer in spirit to a one-orbital Hubbard model.  Heavy-fermion physics, if it exists, must then be emergent: the itinerant carrier of the low-energy theory need not be the bare electron.

The treatment projected to the active band only leads to the proposal of a \emph{Mott semimetal}
\cite{Ledwith2025,Zhao2025,ZhaoAncilla2025,ZhouWangZhang2026}.  One useful description
is the ancilla formulation of Refs.~\cite{Zhao2025,ZhaoAncilla2025},
where the
physical electron $c$ hybridizes with an ancilla fermion $\psi$,
which is Kondo-coupled to local moments represented by another ancilla fermion $\psi'$. The hybridization
$\Phi(\bm k)c^\dagger(\bm k) \psi(\bm k)$ gives the Mott gap, but it  vanishes linearly at the moir\'e-zone center $\Gamma_M$,
leaving a $\psi$-dominated quadratic band touching.  
Complementary approaches based on Hubbard-I/III approximation or diagrammatic calculations \cite{Ledwith2025,VituriBerg2026,Wei2026,Hu2026,Nosov2026} can describe the charge sector formed by $(c, \psi)$  in the ancilla theory, but they typically assume  freely fluctuating local moments and thus can work only at high temperature. We note that if we want to extend the theory down to lower temperature, we must face the challenge of capturing both the itinerant sector and the local moments sector on equal footing. Therefore the $\psi'$ degrees of freedom in the ancilla theory must be kept in the low energy effective theory. Actually we expect an effective Kondo coupling $J_K$ between the itinerant carrier $\psi$ and the local moments from $\psi'$. One major focus of this paper is to derive this coupling from the microscopic model and then discuss the low temperature phases.

One key observation is that the exotic excitation and momentum-selective Mott gap are due to the \textit{mixed-valence} nature of the Mott state, which was already proposed in Ref.~\cite{Zhao2025} assuming a rung-singlet state for the local moments. In this work, we can now formulate a description of the mixed-valence Mott state for a general symmetric phase of the local moments. We start from the topological heavy fermion model (THFM) with dispersive bands $(c_1,c_2)$ and localized orbital $f$. There is a hybridization $\gamma c_1^\dagger f+h.c.$ which gaps out $c_1$ and sets the band gap to the remote band.  We will formulate a theory at general $\gamma/U$  and eventually take the large $\gamma/U$ limit.  In the $\gamma\ll U$ limit, the $f$ orbital can be in a conventional Mott insulator with a fixed valence $n^f_i=\bar n$, with $\bar n$ the integer filling $\bar n=4+\nu$. However, with $\gamma \gtrsim U$, there must be a finite density of doublon and holon excitations for the $f$ orbital in the ground state. Therefore, in the low energy theory, we must keep the doublon $d_\alpha$ and the holon $\tilde h_\alpha$ as elementary degrees of freedom and formulate a theory for the charge sector described by $(c_1, c_2)$ and $(d, \tilde h)$ together. With the linear combinations $f_0\equiv P_GfP_G= \frac{1}{\sqrt{2}} (d+\tilde h)$ and $\psi = \frac{1}{\sqrt2}(d-\tilde h)$, the charge sector is described by $(c_1,c_2, f_0, \psi)$ and we reproduce the ancilla theory \cite{ZhaoAncilla2025}. Now we can interpret $\psi$ as a specific linear combination of the doublon and holon operator such that it is orthogonal to the microscopic electron operator $f$, and thus we dub it \textit{orthogonal fermion}. Note $\psi$ was interpreted as a composite excitation with an electron bound to a particle--hole pair \cite{ZhaoAncilla2025,Ledwith2025}. If we restrict to the Hilbert space with $n^f_i=\bar n, \bar n \pm 1$ by the projection operator $P_G$, we can write $\psi_i \sim P_G \{ f, \delta n^f_i\} P_G$.  However, that is just one way to capture the sign structure of the linear superposition and it is misleading to really view $\psi$ as a bound state.  Actually in the strong mixed-valence regime, a better microscopic operator is $\psi_{0,i}\sim (\delta \hat n^f_i+\tfrac12)^{-1}f_i$.

In the projective limit $\gamma\gg U$, $f$ is gapped out at $\bm k=0$ and $\psi$ dominates at low energy. Therefore, the correct itinerant carrier of the mixed-valence Mott state is the orthogonal fermion instead of the bare electron. In our formulation, we also discover an effective  Kondo coupling $J_K$ between $\psi$ and the local moments.   We have $J_K=\zeta U$ with $\zeta= \frac{\gamma^2}{\gamma^2+U^2/4}$. The full theory is therefore exactly the ancilla theory with $(c, \psi)$ forming the charge sector and Kondo coupled to the local moments represented by $\psi'$. We need to emphasize that the ancilla framework is really necessary.  Due to the mixed-valence nature, the true density of the local moment is actually less than one. Hence it is not obvious at all that a Kondo lattice model is possible, as the Kondo model assumes one spin moment at each site without defects. The ancilla framework naturally solves this issue as $n_\psi$ and $n_{\psi'}$ are constrained to be $N_f-\bar n$ and $\bar n$. As we see later, this gives the correct Luttinger volume in the Kondo screened phase.

We then discuss the consequence of the Kondo coupling. First, from simple $O(J_K^2)$ perturbation calculation, we can recover the scattering rate for charge excitation obtained in recent works based on more complicated diagram calculations \cite{VituriBerg2026,Wei2026,Hu2026,Nosov2026}. To the best of our knowledge, our theory offers the first transparent physical interpretation of the scattering as arising from the Kondo coupling between an emergent charge carrier dominated by  the orthogonal fermion $\psi$ and the local moments.  More importantly, we can now discuss physics at lower temperature where the perturbative calculations break down.  The dimensionless Kondo coupling is $g\equiv N_fJ_K\rho_\psi(0)=2\zeta N_fs^2|M|/U$ at charge neutrality $\bar n=\frac{N_f}{2}$, with flavor degeneracy
$N_f=8$ and  density of states per flavor
$\rho_\psi(0)=2s^2|M|/U^2$, where $s^2=\pi\kstar^2/\Omega_M$ is the
patch-area fraction ($\Omega_M$ is the moir\'e Brillouin-zone
area) and $M$ is the half bandwidth.   In the projective limit
$\zeta=\frac{J_K}{U} \to1$ and $g=2N_fs^2|M|/U$.  Despite $s^2\simeq0.065$ at the magic angle ($\theta=1.081^\circ$),
$2 N_fs^2\simeq 1$ is not a small parameter.   Increasing $|M|/U$ drives $g$ toward
unity, invalidating the weak-coupling Kondo expansion and suggesting a Kondo screened phase instead. It is known that the Kondo coupling flows to strong coupling under the renormalization group (RG) below a temperature scale $T_K \sim \Lambda e^{-\frac{1}{g}}=U e^{-\frac{U}{2N_f s^2 |M|}}$. From self-consistent mean field theory, we find $T_K$ reaches $26$~K at $\theta=1.15^\circ$ already. Inside the Kondo screened phase, we have a heavy semimetal at charge neutrality. For $\nu=-1, -2$, there is already a small gap in the Mott state and increasing $|M|/U$ first closes the gap and then drives the system into a heavy Fermi liquid.  Finally, the anti-Hund's interaction $J_A$ pairs the moments in the
intervalley-singlet channel
\cite{ZhaoZhangRVB2025}.  With neutral pairing amplitude
$\Delta=J_A\langle\psi'\psi'\rangle$, the Kondo condensate transfers
pairing to the charge channel and leads to superconductivity, similar to the recent resonating-valence-bond (RVB) mechanism of Ref.~\cite{ZhaoZhangRVB2025}.

\emph{Mixed valence Mott state and  orthogonal fermion ---} The THFM of TBG \cite{SongBernevig2022} consists of a local orbital $f_{i\alpha}$ and two itinerant orbitals $c_{1;\alpha}(\bm k)$ and $c_{2;\alpha}(\bm k)$. Here $\alpha=(a,\tau, s)$ labels the $N_f=8$ flavors
 combining orbital, valley and spin. $a=\pm$ labels $p\pm ip$ orbitals for $f$. $\tau=K,K'$ and $s=\uparrow, \downarrow$ label the valley and spin. The model is in the form
\begin{align}
H={}&H_{c_1c_2}
+\sum_{\bm k}\bigl(\gamma\,f^{\dagger}_{\bm k}c_{1\bm k}+\mathrm{H.c.}\bigr)
+\frac{U}{2}\sum_i\bigl(n^f_i-4\bigr)^2 ,
\notag\\
H_{c_1c_2}\equiv{}&\sum_{\bm k}\bigl(c_{1\bm k}^{\dagger}\xi_{\bm k}c_{2\bm k}
+\mathrm{H.c.}\bigr)
+M\sum_{\bm k}c_{2\bm k}^{\dagger}\sigma_xc_{2\bm k},
\label{eq:THFMc}
\end{align}
with $\xi_{\bm k}=v_*\tau_z(k_x\sigma_0+\ii k_y\sigma_z)$.  Here $\sigma_\mu$ labels Pauli matrices in the orbital subspace.
$|M|$ sets the active bandwidth and $\gamma$ sets the band gap from the active band to the remote band.   At $\gamma=0$, $(c_1, c_2)$ forms four bands with a quadratic band touching at $E=0$, the same as the low-energy effective model of AB-stacked bilayer graphene.  The $\gamma$ coupling causes a band inversion and then the active flat band is dominated by $c_2$, instead of $f$ at $\bm k=0$.  In the region $|k|>\kstar$, with
$\kstar=\gamma/v_*$ at $U=0$, the low-energy flat band is dominated by $f$.

\begin{figure}[!tp]
\includegraphics[width=0.92\columnwidth]{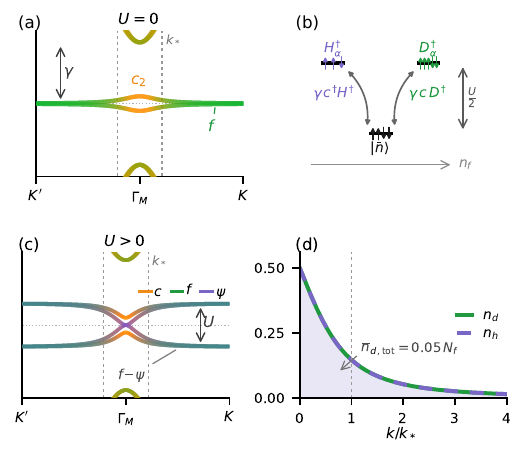}\\[1pt]
\includegraphics[width=0.98\columnwidth]{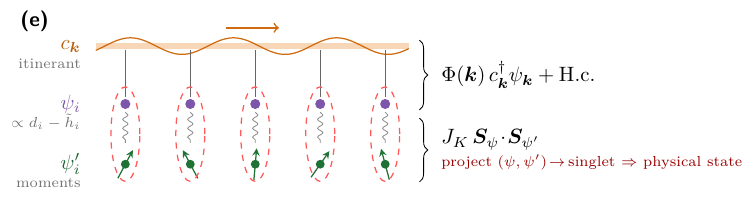}
\caption{
(a) Free fermion bands at $U=0$ along $K'\!-\!\Gamma_M\!-\!K$, coloured by
orbital content.  $\gamma$ inverts the active band: $c_2$-like at
$\Gamma_M$, $f$-like in the regime $|\bm k|>\kstar$.  
(b) Starting from a true Mott insulator of $f$ orbital at a fixed valence $n^f_i=\bar n$, the $\gamma$ hybridization generates a finite density of holon and doublon excitations.  
(c) The bands in the charge sector in the mixed-valence Mott state. At low energy, there is a quadratic band touching dominated by the orthogonal fermion $\psi$.
(d) The momentum distribution of the doublon and holon 
$n_d(k)=n_h(k)=\tfrac12\bigl[1-k/\sqrt{k^2+\kstar^2}\bigr]$ 
per flavor.  
(e) The ancilla framework to capture the emergent heavy fermion physics.  The band $c_{\bm k}$
hybridizes with   $\psi\propto d-\tilde h$ via
$\Phi(\bm k)$, which is further Kondo-coupled to the spinon $\psi'$ by
$J_K\sim U$. 
Projecting $(\psi_i,\psi'_i)$ onto on-site singlets gives the physical state. We interpret $\psi$ and $\psi'$ as the orthogonal fermion and the spinon. In this framework naturally we have the constraints $n_{\psi}=N_f-\bar n$ and $n_{\psi'}=\bar n$.}
\label{fig:schematic}
\end{figure}

We focus on integer fillings $\bar n=4+\nu$. We expect $\gamma\gtrsim U/2$ and hence the $f$ orbital cannot be in a fixed valence $n_{i;f}=\bar n$. In the following we show that there can be a mixed-valence Mott state with unconventional excitations in this regime. Let us start from the standard Mott insulator for the $f$ orbital at filling $\bar n$ with $\gamma=0$.  Then it is known that a Mott gap $\Delta_{\mathrm{M}}=U$ separates the upper Hubbard band (UHB) and the lower Hubbard band (LHB), which represent the doublon and holon excitations. At $\gamma\ll U$ we could treat doublons and holons as
\emph{virtual} excitations, integrate them out, and recover the
familiar Kondo coupling by Schrieffer--Wolff elimination.  At
$\gamma\gtrsim U/2$ they must instead be kept as part of the low-energy
theory, and they acquire a finite density in the ground state
[Fig.~\ref{fig:schematic}(d)].

Restrict the valence to $n_f\in\{\bar n-1,\bar n,\bar n+1\}$ by a
projector $P_G$, so that $|\bar n\rangle$ is the vacuum of the charge
sector.  With
$P_{i;n}$ the projector onto valence $n^f_i=n$, the holon and doublon excitations are created by
$H^{\dagger}_{i\alpha}=P_{i;\bar n-1}f_{i\alpha}P_{i;\bar n}$ and
$D^{\dagger}_{i\alpha}=P_{i;\bar n+1}f^{\dagger}_{i\alpha}P_{i;\bar n}$.     We normalize $d_{i\alpha}=\sqrt{2}\,D_{i\alpha}$ and $\tilde h_{i\alpha}=\sqrt{2}\,H^\dagger_{i\alpha}$ at $\nu=0$. As shown in the supplementary, $d$ and $\tilde h$ are canonical fermions on average when there is no flavor ordering.  Each doublon or holon costs $U/2$, so the Hubbard term is $H_U=\tfrac{U}{2}\sum_i(n_{i;d}-n_{i;\tilde h})$ up to a constant, with $n_{i;d}=\sum_\alpha d^{\dagger}_{i\alpha}d_{i\alpha}$ and $n_{i;\tilde h}=\sum_\alpha\tilde h^{\dagger}_{i\alpha}\tilde h_{i\alpha}$.  The charge sector is described by a Gaussian theory with $c_1, c_2$ and $d, \tilde h$:
\begin{align}
H_0={}&H_{c_1c_2}+\frac{U}{2}\sum_{i}\bigl(n_{i;d}-n_{i;\tilde h}\bigr)
\notag\\
&+\frac{\gamma}{\sqrt2}\sum_{\bm k\alpha}
\bigl[c^{\dagger}_{1\bm k\alpha}\bigl(\tilde h_{\bm k\alpha}+d_{\bm k\alpha}\bigr)
+\mathrm{H.c.}\bigr].
\label{eq:H0}
\end{align}
We can also define linear combinations
$f_{0,i\alpha}=\frac{\tilde h_{i\alpha}+d_{i\alpha}}{\sqrt2}$ and
$\psi_{i\alpha}=\frac{d_{i\alpha}-\tilde h_{i\alpha}}{\sqrt2}$. $\psi$ is 
the \textit{orthogonal fermion}  with no overlap with the
microscopic electron.  The theory is then exactly the ancilla theory
of the THFM \cite{ZhaoAncilla2025}: $\gamma$ couples $c_1$ to the
physical electron alone and $U$ enters as
$(U/2)(f_0^{\dagger}\psi+\mathrm{H.c.})$, with $\psi$ the ancilla
fermion.  Microscopically $\psi=P_G\{f,\delta n\}P_G$ \cite{Zhao2025,Ledwith2025}, but it is better thought of as an orthogonal coherent superposition of the doublon and holon than as a bound state of an electron and a particle--hole pair.  When the mixed-valence fluctuation is strong, the better microscopic operator turns out to be $\psi_0\propto (\delta n+\tfrac12)^{-1}f$ rather than $\{f,\delta n\}$.

We can easily diagonalize $H_0$. When $\gamma\gtrsim U$, $c_1$ and $f$ gap out each other at $\bm k=0$, leaving $c_2, \psi$ at low energy within $|k|<k_*$. In the region $|k|>k_*$, the low energy is dominated by the usual Hubbard band represented by $d, \tilde h$ or $f\pm \psi$.    The effective Hamiltonian projected to the active band is
\begin{equation}
H_{\rm ch}=\sum_{\bm k}\Bigl[M\,F(k)\,c^{\dagger}_{\bm k}\sigma_xc_{\bm k}
+\frac{\Phi(k)}{k}\bigl(c^{\dagger}_{\bm k}(k_x\sigma_0-\ii k_y\sigma_z)
\chi_{\bm k}+\mathrm{H.c.}\bigr)\Bigr],
\label{eq:Hch}
\end{equation}
with $F(k)=\kstar^2/(\kstar^2+k^2)$ the active-band form factor and
$\Phi(k)=\frac{U}{2}\frac{k}{\sqrt{k^2+\kstar^2}}$ the
hybridization amplitude: it rises with slope
$\Phi_\Gamma=\frac{U}{2\kstar}$,
$\kstar=\frac{\sqrt{\gamma^2+U^2/4}}{v_*}$, and saturates at
$U/2$.    Here $c(\bm k)$ is the active
band electron operator and, at $\bm k=0$,
$\chi=\frac{\gamma\psi-(U/2)c_1}{\sqrt{\gamma^2+U^2/4}}$.  One may notice that Eq.~\eqref{eq:Hch} is in
the same form as $H_{c_1c_2}$, but $\chi$ now evolves continuously
from $c_1$ to $\psi$ as $\gamma/U$ grows: in the large $\gamma/U$
limit the quadratic band touching is dominated by the orthogonal
fermion emerging within the $f$ orbital
[Fig.~\ref{fig:schematic}(c)].

\textit{Kondo coupling.---} The above Hamiltonian for the charge sector does not touch the
spin moments within the $n_{i;f}=\bar n$ subspace.  The next natural
question is whether the itinerant carriers $\chi$ can couple to them.
For $\gamma\ll U$ the answer is standard: eliminating the virtual
doublon and holon leaves a Kondo coupling of $c_1$ to the moments,
$\kappa\sum_ic^{\dagger}_{1i}S_ic_{1i}$ with $\kappa=4\gamma^2/U$ and
$S_i^{\alpha\beta}$ as the SU($N_f$) moment of the multiplet.  Projecting
$c_1$ onto the low-energy band $\chi$ leads to
\begin{equation}
H_K=J_K\sum_{\bm k\bm k'}F(k)F(k')
\sum_{\alpha\beta}
\chi^{\dagger}_{\bm k\alpha}S^{\alpha\beta}_{\bm k-\bm k'}
\chi_{\bm k'\beta},
\label{eq:HK}
\end{equation}
with $S^{\alpha\beta}_{\bm q}=N_M^{-1}\sum_ie^{-i\bm q\cdot\bm R_i}
S_i^{\alpha\beta}$.  Here
$J_K=\kappa\alpha^2=\gamma^2U/(\gamma^2+U^2/4)$, with $\alpha=\frac{U/2}{\sqrt{\gamma^2+U^2/4}}$ as the $c_1$ weight of
$\chi$ at $\bm k=0$ and the $k$ dependence of the weight is encoded in the form factor
$F(k)=\kstar^2/(\kstar^2+k^2)$.
Interestingly, taking the large $\gamma/U$ limit of this expression
naively gives $J_K=U$ in the projective limit. In the supplementary we argue that this value of $J_K$ is valid even in the $\gamma\gg U$ region. Note that in the projective limit $J_K$ is expected to be at the scale of $U$.  

Strictly speaking, the density of the local moment is less than one due to the mixed-valence nature of the Mott state. Thus it is not clear whether a Kondo model is well defined when both the charge carriers and the local moments are from the same $f$ orbital.  The most precise framework is the ancilla theory where $\psi$ and $\psi'$ are introduced as auxiliary fermions and projected onto an $SU(N_f)$ spin singlet in the end to get the physical state [Fig.~\ref{fig:schematic}(e)]. Physically $\psi$ represents the orthogonal fermion and $\psi'$ is the neutral spinon. But in this framework naturally we have the constraints $n_{\psi'}=\bar n$ and $n_\psi=N_f-\bar n$. Therefore the low energy effective theory we are going to use is $H_{\rm eff}=H_{\rm ch}+H_K$, with $\chi=\psi$ in the projective limit and separate chemical potentials $\mu_\psi, \mu_{\psi'}$ introduced to fix the two density constraints.  In the following our numerical results evaluate $J_K=\zeta U$ and $F(k)$ at the physical $\gamma(\theta)/U$ with $U=30$ meV fixed just as an illustration. $\gamma$ and other THFM parameters are quoted from the THF parametrization of Ref.~\cite{Calugaru2023}.

\emph{ Heavy semimetal at neutrality.---} The charge sector hosts a
quadratic band touching by $\psi$,
\begin{equation}
H_\psi=-\frac{\Phi_\Gamma^2}{M}\sum_{\bm k}\psi^{\dagger}_{\bm k}
\bigl[(k_x^2-k_y^2)\sigma_x-2k_xk_y\sigma_y\bigr]\psi_{\bm k}.
\label{eq:HQBT}
\end{equation}

Then we consider the Kondo coupling to the local moments. In the free moment regime at $T\gg T_K$, a simple $O(J_K^2)$ calculation gives the scattering rate 
$\Gamma_\psi^{\rm QBT}=\tfrac{N_f+1}{2}\pi s^2\zeta^2|M|$, where
$\zeta\equiv J_K/U=\frac{\gamma^2}{\gamma^2+U^2/4}$
($\zeta\to1$ in the projective limit) and
$s^2=\pi\kstar^2/\Omega_M$ is the patch fraction, reproducing the recent controlled lifetime results for
the THFM \cite{VituriBerg2026,Wei2026,Hu2026,Nosov2026} and
identifying them as Kondo scattering between the emergent orthogonal fermion and the local moments.  The Kondo interpretation not only makes the physical picture transparent, but also reveals the limitation of the perturbative calculations.  It is known that the Kondo coupling will eventually flow to strong coupling at a temperature scale defined as $T_K\sim U e^{-\frac{1}{g}}$, where the dimensionless coupling is 
$g\equiv N_fJ_K\rho_\psi(0)=2\zeta N_fs^2|M|/U$.  It becomes strong once $|M|/U$ is appreciable, with
the full flavor enhancement $N_f$.  The competing ordering scale of the unscreened moments is
ferromagnetic, $T_{\rm FM}\sim Us^2$, with no $N_f$ factor. Therefore we expect a Kondo screened phase away from the magic angle at lower temperature.
  Increasing $|M|$ then drives a
bandwidth-tuned Mott transition of TBG at neutrality in the framework of  the Kondo screening
transition, in the spirit of Kondo-breakdown
criticality \cite{Doniach1977,SiSteglich2010,Lohneysen2007}.

\begin{figure}[t]
\includegraphics[width=\columnwidth]{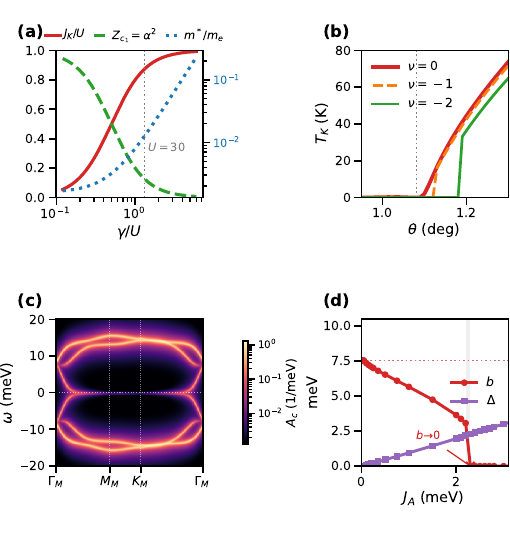}
\caption{
(a) The \(\gamma/U\) crossover at \(\theta=1.15^\circ\) for the charge sector at $\nu=0$, with
\(\gamma,M,v_*\) fixed at their THFM values and \(U\) varied.
Red: \(J_K/U=\gamma^2/(\gamma^2+U^2/4)\).  Green dashed: the
\(c_1\) spectral weight \(Z_{c_1}=\alpha^2\).  The two cross at
\(\gamma=U/2\).  Blue dotted (right axis): the effective mass
of the quadratic band touching,
\(m^*=\hbar^2|M|/(2\alpha^2v_*^2)
=\frac{\hbar^2|M|}{2v_*^2}\bigl(1+\tfrac{4\gamma^2}{U^2}\bigr)\).  The
physical point \(U=30\)~meV sits on the heavy, small-\(Z\) side.
(b) Kondo screening temperature $T_K$ versus twist angle at the 
integer fillings $\nu=0, -1, -2$.    Dotted line: the magic angle, defined by \(M=0\)
(\(\theta=1.081^\circ\)).
(c) Spectral function \(A_c(\omega,\bm k)\) along
\(\Gamma_M\)--\(M_M\)--\(K_M\)--\(\Gamma_M\) at
\(\theta=1.15^\circ\), \(\nu=0\), in the Kondo screened heavy semimetal phase.  The condensate hybridizes the spinon $\psi'$
 with the charge sector and leads to  a heavy band at the Fermi level. 
(d) The mean-field solution \((b,\Delta)\) of
Eq.~\eqref{eq:HbDelta} at \(\theta=1.15^\circ\),
\(\nu=0\), versus the anti-Hund's coupling
\(J_A\).  There is a region with coexisting $b$ and $\Delta$, leading to a superconducting phase.}
\label{fig:overview}
\end{figure}

Figure~\ref{fig:overview}(b) shows the resulting $T_K(\theta)$
from the projected $(c,\psi,\psi')$ mean field:
$T_K$ is exactly zero at
the magic angle $\theta=1.081^\circ$  and rapidly increases to $26$ K already at 
 $1.15^\circ$.  At
the other integer fillings the same transition is at a larger twist angle since increasing $M$ first closes the small gap of the correlated insulator before driving the system into the Kondo screened phase. Below $T_K$ the singlet condensate generates a hybridization
between $\psi$ and $\psi'$, with amplitude $b(k)=b\,F(k)$ and
$b=J_K\frac{1}{N_M}\sum_{\bm k}F(k)\langle\psi'^{\dagger}\psi_{\bm k}\rangle$
(see the supplementary); it turns the spinon $\psi'$ into a
heavy quasi-particle, as shown in Fig.~\ref{fig:overview}(c).  One
can see a new band with small quasi-particle residue $Z$ close to
$E=0$ in the $|\bm k|>\kstar$ region, along with the UHB and LHB.
With $T_K$ increasing at larger $\theta$, this heavy semimetal
crosses over to the free-fermion semimetal. 

\begin{figure}[t]
\includegraphics[width=\columnwidth]{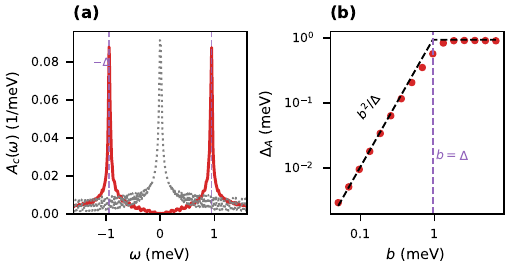}
\caption{
(a) Momentum-integrated \(A_c(\omega)\) inside the superconducting phase, at the \(\theta=1.15^\circ\), \(\nu=0\), \(J_A=1\)~meV solution of Fig.~\ref{fig:overview}(d) (\(b=5.7\)~meV, \(\Delta=0.95\)~meV).  The four point nodes of the
\(|\cos2\phi|\) gap (\(\phi\) is the polar angle of \(\bm k\), measured from the nematic axis) give a V shape, with coherence peaks at
\(\pm\Delta\).  Dotted: the same spectral function $A_c(\omega)$ for the normal state with $\Delta=0$. 
(b) Antinodal gap \(\Delta_A\) versus \(b\) at fixed
\(\Delta=0.95\)~meV (\(\theta=1.15^\circ\), \(\nu=0\)),
crossing over from the proximity form
\(b^2/\Delta\) (dashed) to the locked value \(\Delta\). In the regime $b<\Delta$, we expect a two-gap structure.}
\label{fig:scspec}
\end{figure}

\emph{Nematic nodal superconductivity from RVB mechanism.---} The emergent heavy-fermion picture also makes it possible to discuss a potential superconductivity after the Mott transition through the RVB mechanism. Basically, if the spinon $\psi'$ already has pairing in the Mott state, the condensation of $b$ will transfer the pairing to the itinerant electrons and result in a superconducting phase. As an illustration, we introduce an on-site anti-Hund's coupling $J_A$ from optical phonon
\cite{WuMacDonaldMartin2018,BlasonFabrizio2022,WangPhononTHF2025}. Then the spinon can acquire a pairing term in the mean field:
$\Delta_{ab}=J_A\langle\psi'^{\rm T}_a(\ii s_y\tau_x)\psi'_b\rangle$
\cite{ZhaoZhangRVB2025}, an intervalley ($\tau_x$) spin-singlet
($\ii s_y$) pair with orbital structure $ab$.  Together with the
Kondo condensate, the mean field adds
\begin{equation}
H_{b,\Delta}=b\sum_{\bm k\alpha}F(k)\,
\psi^{\dagger}_{\bm k\alpha}\psi'_{\bm k\alpha}
+\sum_{i,ab}\Delta_{ab}\,
\psi'^{\dagger}_{ia}\,\ii s_y\tau_x\,\psi'^{\dagger}_{ib}
+\mathrm{H.c.},
\label{eq:HbDelta}
\end{equation}
at cost \(b^2/J_K+\sum_{ab}|\Delta_{ab}|^2/J_A\) per site;
\(b\) and \(\Delta_{ab}\) are fixed by minimizing the free
energy. Depending on whether it favors the inter-orbital or intra-orbital pairing, we get either an $s$-wave gapped superconductor or a nematic nodal superconductor. Interestingly, even at $\nu=0$, the pairing of the quadratic band touching of $\psi$ can have nodes, and we find a V shape in the density of states $A_c (\omega)$ [Fig.~\ref{fig:scspec}(a)].  Note that the superconductor emerges when $b$ and $\Delta$ coexist.  Interestingly, there are two regimes.  (I) When $b<\Delta$, the resulting pairing gap is $\Delta_{SC}\sim \frac{b^2}{\Delta}$ and we expect a two-gap structure as in Ref.~\cite{ZhaoZhangRVB2025}. (II) When $b>\Delta$, there is only one gap $\Delta_{SC}=\Delta$. At $\nu=0$ (or $\nu=-2$), there is only a first-order jump of $b$ when decreasing $J_A$, so we only find the regime $b>\Delta$ in our mean field for larger twist angle $\theta$.  At the magic angle, $b=0$ at integer filling and it may gradually increase with doping $x$; the superconductor at $\nu=-2-x$ near the magic angle is therefore likely in the regime $b<\Delta$. Then we expect a two-gap structure, as observed in experiment \cite{ParkNodal2026}.  We note that the superconductor at $\nu=-2-x$ should be in the same phase as the superconductor discussed here at integer filling. Therefore our current emergent heavy-fermion picture may provide a unified framework for the RVB mechanism of superconductivity in TBG.

\emph{Summary.---}  We have derived the ancilla theory for the mixed-valence Mott state of TBG at integer filling through the microscopic model. The phase hosts both itinerant carriers and local moments from the $f$ orbital, and it is quite challenging to incorporate both of them simultaneously in one unified theory. In our framework, we have a charge sector formed by hybridization of the electron $c(\bm k)$ and an emergent orthogonal fermion $\psi(\bm k)$, which further couples to the local moments represented by $\psi'$ through a Kondo coupling $J_K$. At charge neutrality, the Kondo coupling between the quadratic band touching semimetal formed by $\psi$ and the moments $\psi'$ drives the system into a Kondo screened heavy semimetal when increasing the twist angle. If the spinon $\psi'$ already has a pairing in the Mott state, for example from the anti-Hund's coupling, we obtain a superconducting phase after Kondo screening.  To the best of our knowledge, the ancilla framework is the first theory that can describe emergent heavy fermion physics within a one-band model. The same framework has also been applied to the hole-doped cuprates to understand the pseudogap metal and its transition to superconductivity \cite{ZhangSachdev2020}. Therefore we anticipate a unification of the physics in TBG and in the hole-doped cuprates.

\textit{Acknowledgement.---} I thank Taige Wang, Jing-Yu Zhao and Boran Zhou for previous collaborations and discussions.
The work is supported by the Alfred P. Sloan Foundation through a
Sloan Research Fellowship. I have used ChatGPT Sol 5.6 extra high and Claude Fable 5 max for assisting derivations and calculations.

\bibliographystyle{apsrev4-2}
\bibliography{paperRefs}

\appendix
\setcounter{secnumdepth}{3}
\onecolumngrid

\section{Microscopic derivation of the ancilla theory for the topological heavy fermion model}
\label{app:ancilla}

\subsection{Charge sector with holon and doublon operator}
\label{app:ops}

We derive the ancilla-fermion representation used in the main text
directly from the restricted-valence description of the $f$ orbital in the THFM.  As the only interaction in Eq.~(1) of the main text is the Hubbard $U$ on the $f$ orbital, we can first solve the Hubbard $U$ term for one $f$ atom and then discuss its hybridization with the itinerant $c_1$ band. 

The local \(f\) multiplet at moir\'e site \(i\) spans \(N_f=8\)
flavors \(\alpha=(a,\tau,s)\).  We write the charging energy about the
reference valence \(\bar n=\tfrac{N_f}{2}+\nu\) set by the filling,
\begin{equation}
H_U=\frac{U}{2}\sum_i\bigl(n^f_i-\bar n\bigr)^2 ,
\label{eq:HUnbar}
\end{equation}
which at charge neutrality is the particle--hole symmetric
\(\tfrac{U}{2}(n^f_i-\tfrac{N_f}{2})^2\); how the reference should be
chosen at \(\nu\neq0\), and how much it matters, is the subject of
Sec.~\ref{sss:nonzero}.  Everything in this subsection holds for any
integer \(\bar n\).

At low energy we keep the valence \(n_f\in\{\bar n-1,\bar n,\bar n+1\}\),
i.e.\ the \(\binom{N_f}{\bar n-1}+\binom{N_f}{\bar n}
+\binom{N_f}{\bar n+1}\) local states closest to \(\bar n\)---at
neutrality \(56+70+56\).  Both the \(n^f_i=\bar n+1\) and
\(n^f_i=\bar n-1\) excitations cost \(\tfrac{U}{2}\) by
Eq.~\eqref{eq:HUnbar}, and we label them doublon and holon excitations
following the terminology usually used for the \(N_f=2\) Hubbard
model.

We define $P_{i;m}$ as the projection operator onto the $n^f_i=m$ subspace of the $f$ orbital at site $i$. Within this restricted space the holon and the doublon are defined as:
\begin{equation}
H^{\dagger}_{i\alpha}=P_{i;\bar n-1}f_{i\alpha}P_{i;\bar n},
\qquad
D^{\dagger}_{i\alpha}=P_{i;\bar n+1}f^{\dagger}_{i\alpha}P_{i;\bar n}.
\label{eq:Hubbardtransitions}
\end{equation}

Note that $H_i$ and $D_i$ do not have a unique vacuum. Instead they annihilate the local moment subspace defined by $P_{i;\bar n}$:  $H_{i;\alpha} P_{i;\bar n}=0$ and $D_{i;\alpha} P_{i;\bar n}=0$. For simplicity we also define a particle-hole transformed operator:
\begin{equation}
\tilde H_{i\alpha}\equiv H^{\dagger}_{i\alpha}
=P_{i;\bar n-1}f_{i\alpha}P_{i;\bar n}.
\label{eq:PHholon}
\end{equation}
Now  \(\tilde H_{i;\alpha}\)
\emph{creates} a holon  and
\(n_{\tilde H}=N_f-n_H\).  Then we can define two different linear combinations of them:
\begin{equation}
P_Gf_{i\alpha}P_G=\tilde H_{i\alpha}+D_{i\alpha},
\qquad
\psi_{i\alpha}\equiv D_{i\alpha}-\tilde H_{i\alpha},
\label{eq:exactbrightdark}
\end{equation}
with \(P_G=\prod_i\bigl(P_{i;\bar n-1}+P_{i;\bar n}
+P_{i;\bar n+1}\bigr)\) as the Gutzwiller projector.  \(\psi_{i\alpha}\) is the
\textit{orthogonal fermion}.  Technically it can be written as
\begin{equation}
\psi_{i\alpha}=P_G\bigl\{f_{i\alpha},\,\delta n_i\bigr\}P_G ,
\qquad
\delta n_i\equiv n^f_i-\bar n .
\label{eq:psianticomm}
\end{equation}
But it should not be understood as a composite bound state.    The composite three-fermion form
is just a way of writing the \emph{relative} sign of the two
linear combinations, not a binding of \(f\) to a particle--hole pair.

\emph{Canonical algebra.}  The vacuum of the doublon \(D_i\) and the
holon \(H_i\) is the local moment subspace, which has \(\binom{N_f}{\bar
n}\) states \(\ket{\bar n;s}\) with \(s\) the flavor
configuration.  Therefore we cannot use them as standard canonical
fermionic operators to build an effective theory.  

For any operator \(\mathcal O\) we use the notation \(\mathcal O|_{P_{i;\bar n}}\equiv P_{i;\bar
n}\,\mathcal O\,P_{i;\bar n}\), and define a bilinear
\begin{equation}
X_i^{\alpha\beta}\equiv
P_{i;\bar n}\,f^{\dagger}_{i\alpha}f_{i\beta}\,P_{i;\bar n} .
\label{eq:Xdef}
\end{equation}

For each fixed pair
\((\alpha,\beta)\), \(X_i^{\alpha\beta}\) is a single \(\mathcal
M\times\mathcal M\) matrix with \(\mathcal M\equiv\binom{N_f}{\bar n}\) as the dimension of the subspace $P_{i;\bar n}$.  We write \(\identity_{\mathcal M}\) for
the identity there. We can then define the spin operator from the traceless part:
\begin{equation}
S_i^{\alpha\beta}\equiv X_i^{\beta\alpha}
-\frac{\bar n}{N_f}\,\delta_{\alpha\beta}\,\identity_{\mathcal M}
=P_{i;\bar n}f^{\dagger}_{i\beta}f_{i\alpha}P_{i;\bar n}
-\frac{\bar n}{N_f}\,\delta_{\alpha\beta}\,\identity_{\mathcal M} ,
\qquad
\sum_\alpha S_i^{\alpha\alpha}=0 ,
\label{eq:Xsplit}
\end{equation}
where each side is an \(\mathcal M\times\mathcal M\) matrix.  For \(N_f=2\)  at half
filling, \(S_i^{\alpha\beta}=(\bm S_i\cdot\bm\sigma)^{\alpha\beta}\)
with \(\bm S_i\) the ordinary spin-\(\tfrac12\) operator and $\sigma$ the Pauli matrix.   For general $N_f$, \(S_i\)  can be also defined through
the traceless generators \(T^a\) of \(\mathrm{SU}(N_f)\),
\begin{equation}
S_i^a=\sum_{\alpha\beta}f^{\dagger}_{i\alpha}T^a_{\alpha\beta}
f_{i\beta}\Big|_{P_{i;\bar n}},
\qquad
S_i^{\alpha\beta}=2\sum_a T^a_{\alpha\beta}S_i^a ,
\label{eq:Sgenerators}
\end{equation}
with \(T^a\) here the \(N_f\times N_f\) matrices of the
\emph{fundamental} representation, normalized
\(\Tr_{N_f}T^aT^b=\tfrac12\delta^{ab}\).  This normalization fixes a
basis of the Lie algebra and does not depend on the multiplet; the
representation dependence sits entirely in the operators \(S_i^a\),
which are the \(\mathcal M\times\mathcal M\) representation matrices
of the rank-\(\bar n\) antisymmetric irrep generated by the
second-quantized bilinear.  Their normalization is \emph{not}
\(\tfrac12\): it is set by the Dynkin index of that irrep,
\(\Tr_{\mathcal M}S_i^aS_i^b
=\tfrac12\binom{N_f-2}{\bar n-1}\delta^{ab}\), i.e.\ \(10\,\delta^{ab}\)
at neutrality (\(N_f=8\), \(\bar n=4\)); equivalently
\(\sum_aS_i^aS_i^a=C_2\,\identity_{\mathcal M}\) with the quadratic
Casimir \(C_2=\bar n(N_f-\bar n)(N_f+1)/(2N_f)=9\).  The resolution
identity \(S_i^{\alpha\beta}=2\sum_aT^a_{\alpha\beta}S_i^a\) is
nevertheless representation independent: it follows from the
completeness relation
\(\sum_aT^a_{\alpha\beta}T^a_{\gamma\delta}
=\tfrac12\bigl(\delta_{\alpha\delta}\delta_{\beta\gamma}
-\tfrac1{N_f}\delta_{\alpha\beta}\delta_{\gamma\delta}\bigr)\) of the
fundamental matrices, with the multiplet entering only through
\(\hat n^f_i=\bar n\) in the trace subtraction of
Eq.~\eqref{eq:Xsplit}.

The algebra of \(\tilde H\) and \(D\) now follows from the identity
\begin{equation}
f_{i\alpha}P_{i;m}=P_{i;m-1}f_{i\alpha},
\qquad
f^{\dagger}_{i\alpha}P_{i;m}=P_{i;m+1}f^{\dagger}_{i\alpha},
\qquad
P_{i;m}P_{i;m'}=\delta_{mm'}P_{i;m} .
\label{eq:intertwine}
\end{equation}
The cross term needs no computation: \(\tilde H_{i\alpha}
D^{\dagger}_{i\beta}\) ends in \(P_{i;\bar n}P_{i;\bar n+1}=0\) and
\(D^{\dagger}_{i\beta}\tilde H_{i\alpha}\) in \(P_{i;\bar n}P_{i;\bar
n-1}=0\), so \(\{\tilde H_{i\alpha},D^{\dagger}_{i\beta}\}=0\)
identically.  On the other hand,
\(\tilde H_{i\alpha}\tilde H^{\dagger}_{i\beta}=P_{i;\bar
n-1}f_{i\alpha}f^{\dagger}_{i\beta}P_{i;\bar n-1}\) and \(\tilde
H^{\dagger}_{i\beta}\tilde H_{i\alpha}=P_{i;\bar
n}f^{\dagger}_{i\beta}f_{i\alpha}P_{i;\bar n}\), and the two are
supported in \emph{different} sectors, so the anticommutator is block
diagonal:
\begin{align}
\{\tilde H_{i\alpha},\tilde H^{\dagger}_{i\beta}\}
&=\underbrace{P_{i;\bar n}f^{\dagger}_{i\beta}f_{i\alpha}
P_{i;\bar n}}_{\textstyle X_i^{\beta\alpha}}
+P_{i;\bar n-1}\bigl(\delta_{\alpha\beta}
-f^{\dagger}_{i\beta}f_{i\alpha}\bigr)P_{i;\bar n-1},
\notag\\
\{D_{i\alpha},D^{\dagger}_{i\beta}\}
&=\underbrace{P_{i;\bar n}f_{i\alpha}f^{\dagger}_{i\beta}
P_{i;\bar n}}_{\textstyle\delta_{\alpha\beta}\identity_{\mathcal M}
-X_i^{\beta\alpha}}
+P_{i;\bar n+1}f^{\dagger}_{i\beta}f_{i\alpha}P_{i;\bar n+1} .
\label{eq:blockalgebra}
\end{align}
The second block in each line acts only on states that already carry a
defect; it governs two-defect processes and is suppressed at low
defect density.  The first block is on the local moment subspace. Inserting Eq.~\eqref{eq:Xsplit},
\begin{equation}
\{\tilde H_{i\alpha},\tilde H^{\dagger}_{i\beta}\}\Big|_{P_{i;\bar n}}
=\frac{\bar n}{N_f}\,\delta_{\alpha\beta}\,\identity_{\mathcal M}+S_i^{\alpha\beta},
\qquad
\{D_{i\alpha},D^{\dagger}_{i\beta}\}\Big|_{P_{i;\bar n}}
=\Bigl(1-\frac{\bar n}{N_f}\Bigr)\delta_{\alpha\beta}\,\identity_{\mathcal M}
-S_i^{\alpha\beta},
\qquad
\{\tilde H_{i\alpha},D^{\dagger}_{i\beta}\}=0 .
\label{eq:exactalgebra}
\end{equation}
Thus neither \(\tilde H\) nor \(D\) is a canonical fermion, and the
obstruction is precisely \(S_i\): creating a doublon/holon and annihilating it later can cause a spin flip.  

We can now define \(\tilde h\) and \(d\) for general \(\bar n\):
\begin{equation}
\tilde h_{i\alpha}=\frac{\tilde H_{i\alpha}}{\sqrt{\bar n/N_f}},
\qquad
d_{i\alpha}=\frac{D_{i\alpha}}{\sqrt{1-\bar n/N_f}}.
\label{eq:canonicaltransitions}
\end{equation}
Eq.~\eqref{eq:exactalgebra} then becomes
\begin{align}
\{\tilde h_{i\alpha},\tilde h^{\dagger}_{i\beta}\}
&=\delta_{\alpha\beta}\,\identity_{\mathcal M}
+\frac{N_f}{\bar n}\,S_i^{\alpha\beta},
\notag\\
\{d_{i\alpha},d^{\dagger}_{i\beta}\}
&=\delta_{\alpha\beta}\,\identity_{\mathcal M}
-\frac{N_f}{N_f-\bar n}\,S_i^{\alpha\beta},
\notag\\
\{\tilde h_{i\alpha},d^{\dagger}_{i\beta}\}&=0 ,
\label{eq:deformedalgebra}
\end{align}
and the restricted electron and the orthogonal fermion read
\begin{equation}
f_{0,i\alpha}\equiv P_Gf_{i\alpha}P_G
=\sqrt{\tfrac{\bar n}{N_f}}\;\tilde h_{i\alpha}
+\sqrt{1-\tfrac{\bar n}{N_f}}\;d_{i\alpha},
\qquad
\psi_{i\alpha}
=\sqrt{\tfrac{\bar n}{N_f}}\;d_{i\alpha}
-\sqrt{1-\tfrac{\bar n}{N_f}}\;\tilde h_{i\alpha}.
\label{eq:PGfgeneral}
\end{equation}
 At \(\bar
n/N_f=\tfrac12\),
Eqs.~\eqref{eq:canonicaltransitions}--\eqref{eq:PGfgeneral} reduce to
\(\tilde h=\sqrt2\,\tilde H\), \(d=\sqrt2\,D\) and
\(f_0,\psi=(\tilde h+d)/\sqrt2,\,(d-\tilde h)/\sqrt2\), the forms used
in the main text for charge neutrality.

\emph{The free-moment ensemble.}  $\tilde h$ and $d$ are still not canonical, so naively we still cannot use them to form an effective theory. This issue can be removed in the local moment regime at high temperature where each local moment subspace at site $i$ is freely fluctuating and  is thus described by the maximally mixed density matrix
\begin{equation}
\rho_{\rm lm}=P_{i;\bar n}\Big/\mathcal M ,
\qquad
\langle\mathcal O\rangle_{\rm lm}
\equiv\Tr\bigl(\rho_{\rm lm}\,\mathcal O\bigr),
\label{eq:freemoment}
\end{equation}
so that all \(\mathcal M\) configurations are equally probable and no
moment order is present.  This is the appropriate reference at
temperatures below \(U\) but above the ordering and Kondo screening temperature scales.

On average,
\(\langle\identity_{\mathcal M}\rangle_{\rm lm}=1\), while
\(\langle S_i\rangle_{\rm lm}=0\), because in this ensemble each
flavor is occupied with probability \(\bar n/N_f\) and off-diagonal
averages vanish by flavor conservation,
\(\langle X_i^{\beta\alpha}\rangle_{\rm lm}
=(\bar n/N_f)\,\delta_{\alpha\beta}\).  Averaging
Eq.~\eqref{eq:deformedalgebra} therefore leads to the standard canonical anti-commutation algebra:
\begin{align}
\bigl\langle\{\tilde h_{i\alpha},\tilde h^{\dagger}_{i\beta}\}
\bigr\rangle_{\rm lm}&=\delta_{\alpha\beta},
\notag\\
\bigl\langle\{d_{i\alpha},d^{\dagger}_{i\beta}\}
\bigr\rangle_{\rm lm}&=\delta_{\alpha\beta},
\notag\\
\bigl\langle\{\tilde h_{i\alpha},d^{\dagger}_{i\beta}\}
\bigr\rangle_{\rm lm}&=0 ,
\label{eq:thermalmetric}
\end{align}

Equation~\eqref{eq:thermalmetric} is the precise sense in which
\(\tilde h\) and \(d\) are canonical fermions: canonical \emph{in the mean over
the free-moment ensemble}, and not as operators.  The
failure is exactly the \(S_i\)
term of Eq.~\eqref{eq:deformedalgebra}.  In particular the
canonical number operator is not the defect projector: on any
single-doublon state,
\(\sum_\alpha d^{\dagger}_{i\alpha}d_{i\alpha}\) evaluates to
\((\bar n+1)/(1-\bar n/N_f)=10\) rather than to one; only its
ensemble average enters the Gaussian theory.  The factor is a norm
of the composite operator, not a correction to observables: to
leading order in the hybridization the Gaussian occupation
\(\langle n_{i;d}\rangle\) equals the physical doublon
probability \(\langle P_{i;\bar n+1}\rangle\), because the
canonicalized amplitude \(\gamma\sqrt{1-\bar n/N_f}\) is the
exact transition matrix element, so the admixed weights coincide.
The channel occupations of Fig.~\ref{fig:schematic}(d) are
therefore leading-order estimates of the physical defect
probabilities, with \(O(1)\) corrections in the mixed-valence
regime. Later we will see that this effect leads to a Kondo coupling between $\psi$ and the local moment.

For now let us just treat $\tilde h$ and $d$ as standard canonical fermions in the free-moment ensemble.  Strictly speaking there is also a hard-core constraint which constrains $n_{i;d}+n_{i;h} \leq 1$, but we are going to ignore it in the dilute limit with $n_d, n_h \ll 1$.  Then we can rewrite
Eq.~\eqref{eq:THFMc} in the main text in terms of $(c_1, c_2, \tilde h,d)$ or $(c_1, c_2, f, \psi)$.  The conduction term
\(H_{c_1c_2}\) contains no \(f\) and is untouched, so only two pieces
need translating.  In the restricted space the valence deviation is
\(n^f_i-\bar n=n_{i;d}-n_{i;h}\), and with at most one defect per site
\((n^f_i-\bar n)^2=n_{i;d}+n_{i;h}\), which under the particle--hole
relabeling \(n_{i;\tilde h}=N_f-n_{i;h}\) turns into
\(n_{i;d}-n_{i;\tilde h}\) up to a constant; the hybridization follows
from Eq.~\eqref{eq:PGfgeneral}.  Hence, at integer filling \(\bar n\),
\begin{equation}
H=H_{c_1c_2}
+\frac{U}{2}\sum_i\bigl(n_{i;d}-n_{i;\tilde h}\bigr)
+\gamma\sum_{\bm k\alpha}\Bigl[
\sqrt{\tfrac{\bar n}{N_f}}\;\tilde h^{\dagger}_{\bm k\alpha}c_{1\bm k\alpha}
+\sqrt{1-\tfrac{\bar n}{N_f}}\;d^{\dagger}_{\bm k\alpha}c_{1\bm k\alpha}
+{\rm H.c.}\Bigr].
\label{eq:Hhd}
\end{equation}
The two excitation channels sit at \(\mp U/2\) and couple to \(c_1\) with
\emph{unequal} amplitudes \(V_{\tilde h}=\gamma\sqrt{\bar n/N_f}\) and
\(V_d=\gamma\sqrt{1-\bar n/N_f}\), whose squares add to \(\gamma^2\).
At \(\bar n/N_f=\tfrac12\) both equal \(\gamma/\sqrt2\) and
Eq.~\eqref{eq:Hhd} is Eq.~\eqref{eq:H0} of the main text.

\emph{The model in the \(f_0,\psi\) basis.}  Inverting
Eq.~\eqref{eq:PGfgeneral},
\(\tilde h=\sqrt{\bar n/N_f}\,f_0-\sqrt{1-\bar n/N_f}\,\psi\) and
\(d=\sqrt{1-\bar n/N_f}\,f_0+\sqrt{\bar n/N_f}\,\psi\), the two
hybridization terms recombine into one: \(V_{\tilde h}\tilde
h^{\dagger}+V_dd^{\dagger}=\gamma f_0^{\dagger}\).  The conduction
electron therefore couples to the restricted physical electron and to
nothing else, while the whole of \(U\) is carried by the
\(f_0\)--\(\psi\) sector,
\begin{equation}
H=H_{c_1c_2}
+\gamma\sum_{\bm k\alpha}\bigl(f^{\dagger}_{0\bm k\alpha}c_{1\bm k\alpha}
+{\rm H.c.}\bigr)
-\Delta_\psi \sum_{i\alpha}\bigl(n_{i;f_0}-n_{i;\psi}\bigr)
+\frac{\bar U}{2}\sum_{i\alpha}\bigl(f^{\dagger}_{0i\alpha}\psi_{i\alpha}
+{\rm H.c.}\bigr),
\label{eq:Hfpsi}
\end{equation}
with the two filling-dependent scales
\begin{equation}
\bar U=2\sqrt{\frac{\bar n}{N_f}\Bigl(1-\frac{\bar n}{N_f}\Bigr)}\;U
=U\sqrt{1-\frac{4\nu^2}{N_f^2}} ,
\qquad
\Delta_\psi=\Bigl(\frac{\bar n}{N_f}-\frac12\Bigr)U
=\frac{\nu}{N_f}\,U ,
\label{eq:Ubardelta}
\end{equation}
using \(\bar n=N_f/2+\nu\).  At neutrality \(\Delta_\psi=0\) and
\(\bar U=U\), so \(U\) enters \emph{only} as the off-diagonal
\((U/2)(f_0^{\dagger}\psi+{\rm H.c.})\), as quoted in the main text. The non-zero integer filling needs some special attention.

\subsubsection{Nonzero integer filling}
\label{sss:nonzero}

At \(\nu\neq0\) there is no particle-hole symmetry, so we need to add a chemical potential $\mu$ to fix the total density of electrons: $-\mu (n_f+n_{c_1}+n_{c_2})$ in  the full model.  In this case we also need to be careful about the Hubbard $U$ term.  Microscopically the Hubbard $U$ on the $f$ orbital should be $H_U=\frac{U}{2} (n^f_i-\frac{N_f}{2})^2$ as our reference density should be the charge neutrality. In practice, to incorporate effects such as the repulsion between $c$ and $f$, one may instead use a renormalized \(U\) with a filling-shifted
reference \cite{LauColeman2025}:
\begin{equation}
H_U=\frac{U}{2}\sum_i\Bigl(n^f_i-\frac{N_f}{2}-\kappa\nu\Bigr)^2 ,
\label{eq:HUkappa}
\end{equation}
which interpolates between the bare microscopic form,
\(\kappa=0\), whose reference is charge neutrality, and
\(\kappa=1\), the reference tracking the filling and assumed in 
Eq.~\eqref{eq:HUnbar}.    Expanding about the actual valence
\(n^f_i=\bar n+n_{i;d}-n_{i;h}\), we obtain
\begin{equation}
H_U=\frac{U}{2}\sum_i\bigl(n_{i;d}+n_{i;h}\bigr)
+\epsilon_f\sum_i\bigl(n_{i;d}-n_{i;h}\bigr)+\text{const},
\qquad
\epsilon_f=U\nu(1-\kappa) .
\label{eq:epsf}
\end{equation}
The change is really from the second term.  Rotating to the
\(f_0,\psi\) basis with \(n_{i;h}=N_f-n_{i;\tilde h}\) and
\(n_d+n_{\tilde h}=n_{f_0}+n_\psi\), Eq.~\eqref{eq:epsf} becomes
\begin{equation}
H_U=-\Delta_\psi\sum_{i\alpha}\bigl(n_{i;f_0}-n_{i;\psi}\bigr)
+\frac{\bar U}{2}\sum_{i\alpha}
\bigl(f^{\dagger}_{0i\alpha}\psi_{i\alpha}+{\rm H.c.}\bigr)
+\epsilon_f\sum_{i\alpha}\bigl(n_{i;f_0}+n_{i;\psi}\bigr)
+\text{const},
\label{eq:epsfpsi}
\end{equation}
which is exactly the \(U\)-dependent part of Eq.~\eqref{eq:Hfpsi} plus the new term \(\epsilon_f\).  The full Hamiltonian, including
the chemical potential, is therefore
\begin{equation}
H=H_{c_1c_2}
+\gamma\sum_{\bm k\alpha}\bigl(f^{\dagger}_{0\bm k\alpha}c_{1\bm k\alpha}
+{\rm H.c.}\bigr)
+H_U-\mu\,n_c ,
\qquad
n_c=n_f+\Bigl(n_{c_1}-\tfrac{N_f}{2}\Bigr)
+\Bigl(n_{c_2}-\tfrac{N_f}{2}\Bigr),
\label{eq:Hnu}
\end{equation}
with \(H_U\) from Eq.~\eqref{eq:epsfpsi} and
\(n_f=n_{f_0}+n_\psi-N_f+\bar n\). $\mu$ is introduced to enforce the constraint that $n_c=4+\nu$. Note that here $n_f \neq n_{f_0}$.  Instead, $n_f-n_{f_0}=\delta n_\psi$ where $\delta n_\psi\equiv n_\psi-(N_f-\bar n)$, with $\bar n=4+\nu$.  We always have $n_{f}=\bar n+n_d-n_h=(\bar n-N_f)+n_d+n_{\tilde h}$.

The above theory resembles the ancilla theory for the charge sector of
Ref.~\cite{ZhaoAncilla2025}.  At charge neutrality they are exactly
the same.  However, at \(\nu\neq0\), there are some crucial differences:

\begin{itemize}
    \item In the ancilla theory, there is no $\epsilon_f=U \nu (1-\kappa)$ term.
    \item In the ancilla theory, we really interpret $f_0$ as the bare $f$ operator and therefore $n_f=n_{f_0}$ and the chemical potential $\mu_c$ is introduced to fix $n_c=n_{f_0}+(n_{c_1}-\frac{N_f}{2})+(n_{c_2}-\frac{N_f}{2})=4+\nu$. 
    \item In the ancilla theory,  $\Delta_\psi$ is not fixed and is self consistently determined to impose $\delta n_\psi=n_\psi-(N_f-\bar n)=0$.
\end{itemize}

Therefore there are two different schemes  to do the calculation for the charge sector at non-zero integer filling.
\emph{Scheme~I (microscopic).}  Keep the $\epsilon_f$ term  and
place no constraint on \(n_\psi\).  The natural variables are then  \(\tilde h,d\) and  so \(n_f=n_d+n_{\tilde h}-N_f+\bar n\).  \(\bar U\) and \(\Delta_\psi\) are fixed by \(U\) and the
filling through Eq.~\eqref{eq:Ubardelta}.  Here \(f_0\) and \(\psi\)
are merely a rotation of \((\tilde h,d)\) and we should not expect \(n_{f_0}=n_f\). The Hamiltonian is
\begin{equation}
H_{\rm I}=H-\mu\,n_c^{\rm I},
\qquad
n_c^{\rm I}=n_{f_0}+\delta n_\psi
+\Bigl(n_{c_1}-\tfrac{N_f}{2}\Bigr)+\Bigl(n_{c_2}-\tfrac{N_f}{2}\Bigr)
=4+\nu ,
\label{eq:schemeI}
\end{equation}
using \(n_f=n_{f_0}+\delta n_\psi\), so that
\(n_c^{\rm I}=n_c^{\rm II}+\delta n_\psi\) and
with \(\Delta_\psi\) fixed by Eq.~\eqref{eq:Ubardelta} and not adjusted.
Generically we have 
\(\delta n_\psi\neq0\) for an arbitrary $\kappa$.

\emph{Scheme~II (ancilla).}  Identify \(n_{f_0}=n_f\) and always fix the constraint $\delta n_\psi=0$.  The Hamiltonian is
\begin{equation}
H_{\rm II}=H-\mu\,n_c^{\rm II}-\lambda_\psi n_\psi ,
\qquad
n_c^{\rm II}=n_{f_0}
+\Bigl(n_{c_1}-\tfrac{N_f}{2}\Bigr)+\Bigl(n_{c_2}-\tfrac{N_f}{2}\Bigr)
=4+\nu ,
\qquad
n_\psi=N_f-\bar n .
\label{eq:schemeII}
\end{equation}
Here \(\lambda_\psi\) shifts
\(\Delta_\psi\) away from Eq.~\eqref{eq:Ubardelta}, and the density of $\psi$ is fixed separately as it is identified as an auxiliary fermion.

\begin{figure}[!htb]
\includegraphics[width=0.82\textwidth]{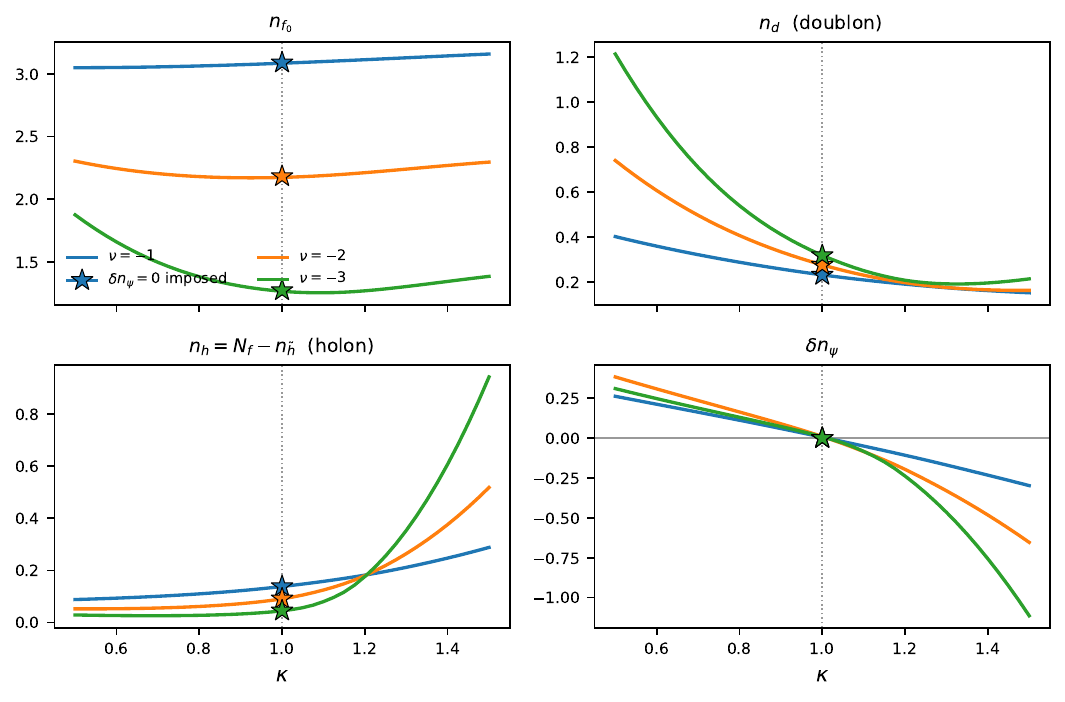}
\caption{Occupations versus \(\kappa\) near the magic angle
(\(\theta=1.05^\circ\): \(\gamma=24.75\,\)meV,
\(|M|=3.69\,\)meV, \(U=30\,\)meV, temperature
\(T=0.02\,\)meV) for \(\nu=-1,-2,-3\), computed from
Eq.~\eqref{eq:schemeI} with \(n_c=4+\nu\) held fixed: \(n_{f_0}\), the
doublon and holon densities \(n_d\) and \(n_h=N_f-n_{\tilde h}\), and
the mismatch \(\delta n_\psi\).  Stars mark the solution of
Eq.~\eqref{eq:schemeII}, which has no \(\kappa\); near \(\kappa=1\)
the two coincide within the linewidth.  \(n_d>n_h\) throughout: the
valence stays mixed in both cases.}
\label{fig:schemes}
\end{figure}

The whole difference is therefore whether \(\lambda_\psi n_\psi\) is a separate chemical potential term or not, or equivalently whether we have $\delta n_\psi=0$.  From direct numerical calculation, we find that they are equivalent at a specific $\kappa_* \approx 1$, as shown in Fig.~\ref{fig:schemes}. If $\kappa$ deviates strongly from $1$, the first scheme is also not justified as $n_d$ or $n_h$ becomes large and away from the dilute limit. Therefore for the effective theory to be valid, we should choose $\kappa \approx 1$ and therefore the two theories are in agreement. In this work we mainly focus on the ancilla theory to deal with non-zero $\nu$. Later we will see that the ancilla theory gives the correct Luttinger Fermi surface for the Kondo screened phase, while the first approach has difficulty dealing with the density of the local moment in the Kondo screened phase.

\subsection{Effective Kondo coupling}
\label{app:exactsigma}

In the previous subsection, we provided an effective theory for the
charge sector while assuming the local moment is freely fluctuating.
The key assumption is the maximally mixed density matrix
\(\rho_{\rm lm}=P_{i;\bar n}/\mathcal M\) of Eq.~\eqref{eq:freemoment}.
Strictly speaking, this is true even at the ground-state level as long
as the ground state is not ordered: for any many-body state invariant
under global \(\mathrm{SU}(N_f)\), the reduced density matrix of site \(i\) commutes with every
on-site generator, and since the \(\bar n\) multiplet is an
\emph{irreducible} representation of \(\mathrm{SU}(N_f)\) (the
rank-\(\bar n\) antisymmetric tensor), Schur's lemma forces its block
in the \(P_{i;\bar n}\) subspace to be proportional to
\(P_{i;\bar n}\).  Then we expect a similar effective theory in terms
of \(\tilde h,d\), or equivalently \(f_0,\psi\).  However, this
effective theory only captures the charge sector.  One natural
question is whether the charge sector couples to the local moments or
not.  Here we will derive a Kondo coupling between the itinerant
carriers and the local moments.

Before we proceed, we point out a subtlety for writing down an effective Kondo model. As we have shown, the number of local moment sites is less than the total number of sites, or equivalently $\langle P_{i;\bar n} \rangle <1$.  Then it is not clear at all that we can put one local spin at each site. The best and maybe the only framework is then to use the ancilla theory where two ancilla fermions $\psi_i$ and $\psi'_i$ are introduced for each moir\'e site and then constrained to be in an $SU(N_f)$ singlet in the end.  With a finite $\Phi(\bm k) c^\dagger(\bm k) \psi(\bm k)$ coupling, we can still interpret $\psi$ as the orthogonal fermion and $\psi'$ as the local moment. Now very naturally we can have $n_{i;\psi'}=\bar n=4+\nu$.  Note that this does not necessarily mean that the physical local moment density is $\bar n$. Due to the mixed-valence nature of the state, we still have defects in the physical state, which should be automatically captured in the non-trivial ancilla framework.  Strictly speaking, one can perform variational wavefunction study and decide the mean field order parameter from minimizing the microscopic free energy. However, this approach is quite challenging and here we take a poor man's approach and try to formulate an effective theory in the $(c, \psi, \psi')$ enlarged Hilbert space.  We already showed that the Hubbard $U$ introduces the coupling $\Phi(\bm k)$ between $c$ and the orthogonal fermion $\psi$. Now with the interpretation of $\psi'$ as the spinon of the local moment, we can derive an effective Kondo coupling and then do the standard mean field calculation. In this framework, $n_{i;\psi'}=\bar n$ and we can have a standard Kondo lattice model.

 Our task is to derive an effective Kondo coupling between the itinerant carriers and the local moments. We assume
a quasi-static local moment configuration \(\ket{s}\in P_{i;\bar n}\) and calculate the
self-energy that \(c_1\) feels.  Then we can extract the effective Kondo coupling by matching the same self energy in the effective theory.  Throughout this subsection we ignore
the corrections from the finite defect density with non-zero $n_d$ and $n_h$.  

We assume $J_{\rm eff}\equiv\max\bigl(T_K,\,J_{\rm RKKY},\,J_A,\ldots\bigr)
\;\ll\;\gamma,\,U $ and  the local moments are frozen on the charge time scales
\(1/\gamma\), \(1/U\): the self-energy can be evaluated
configuration by configuration at fixed \(|s\rangle\).   Write \(x=\bar n/N_f\) and \(u=U/2\).  The starting point
is the Gaussian charge theory already constructed in
Eq.~\eqref{eq:Hhd}: the site block consists of the two levels
\(d\) at \(+u\) and \(\tilde h\) at \(-u\), coupled to
\(c_1\) through the two hybridization channels
\(\gamma\sqrt{x}\,\tilde h^{\dagger}c_1\) and
\(\gamma\sqrt{1-x}\,d^{\dagger}c_1\).  The one change is that we
no longer view \(d\) and \(\tilde h\) as canonical fermions.  At a
frozen moment configuration their exact anticommutators are the
deformed algebra Eq.~\eqref{eq:deformedalgebra}.  This is precisely
the \(S\)-dependence that the free-moment average
Eq.~\eqref{eq:thermalmetric} erased.  With a noncanonical metric it
is not obvious how to diagonalize the theory.  No diagonalization is
needed, however, for the self-energy.  It only requires the
site-block Green's functions, and these follow from the equation of
motion without ever invoking \(H_U\).  In the Gaussian theory the
decoupled site levels evolve trivially, \(d(t)=e^{-\ii ut}d\) and
\(\tilde h(t)=e^{+\ii ut}\tilde h\).  This is the same evolution
the microscopic \(H_U\) generates, since \(d^{\dagger}\) connects
only the \(\bar n\to\bar n+1\) sectors.  For \(a=d,\tilde h\)
with level energy \(\epsilon_a=\pm u\), the retarded function
\(G_{aa^{\dagger}}(t)=-\ii\theta(t)\,\langle
s|\{a(t),a^{\dagger}\}|s\rangle\) obeys
\begin{equation}
(z-\epsilon_a)\,G_{aa^{\dagger}}(z)
=\bigl\langle s\bigl|\{a,a^{\dagger}\}\bigr|s\bigr\rangle :
\label{eq:eomclosure}
\end{equation}
the level structure contributes only the pole position, and
\emph{all} of the nontrivial physics is carried by the equal-time
anticommutator on the right-hand side.  Inserting the deformed algebra
Eq.~\eqref{eq:deformedalgebra} evaluated in the frozen state---write
\(S_i^{\alpha\beta}\big|_s\equiv\langle
s|S_i^{\alpha\beta}|s\rangle\)---
\begin{equation}
G_{\tilde h}^{\alpha\beta}(z)
=\frac{\delta_{\alpha\beta}+S_i^{\alpha\beta}\big|_s/x}{z+u},
\qquad
G_{d}^{\alpha\beta}(z)
=\frac{\delta_{\alpha\beta}-S_i^{\alpha\beta}\big|_s/(1-x)}{z-u},
\label{eq:Gtransition}
\end{equation}
and there is no cross propagator because
\(\{\tilde h,d^{\dagger}\}=0\).  Integrating out the two channels
of Eq.~\eqref{eq:Hhd} gives
\(\Sigma_{c_1}=\gamma^2x\,G_{\tilde h}+\gamma^2(1-x)\,G_d\),
whose numerators recombine into the unnormalized transition residues
\(P_{i;\bar n}f^{\dagger}_{i\beta}f_{i\alpha}P_{i;\bar n}
=x\delta_{\alpha\beta}+S_i^{\alpha\beta}\) and
\(P_{i;\bar n}f_{i\alpha}f^{\dagger}_{i\beta}P_{i;\bar n}
=(1-x)\delta_{\alpha\beta}-S_i^{\alpha\beta}\).  In components,
\begin{align}
\Sigma_{c_1,i}^{\alpha\beta}(\omega)
&=\gamma^2\left[
\frac{\tfrac{\bar n}{N_f}\,\delta_{\alpha\beta}
+S_i^{\alpha\beta}\big|_s}{\omega+U/2}
+\frac{\bigl(1-\tfrac{\bar n}{N_f}\bigr)\delta_{\alpha\beta}
-S_i^{\alpha\beta}\big|_s}{\omega-U/2}
\right]
\notag\\
&=\underbrace{\frac{\gamma^2\,(\omega-\Delta_\psi)}
{\omega^2-U^2/4}}_{\textstyle\bar\Sigma(\omega)}
\,\delta_{\alpha\beta}
\;\underbrace{-\;\frac{\gamma^2U}{\omega^2-U^2/4}\,
S_i^{\alpha\beta}\big|_s}_{\textstyle\delta\Sigma_{i}^{\alpha\beta}(\omega)} ,
\label{eq:configSigmaA}
\end{align}
with \(\Delta_\psi=(\bar n/N_f-\tfrac12)U\) as in
Eq.~\eqref{eq:Ubardelta}; at neutrality \(\Delta_\psi=0\) and the mean
reduces to \(\gamma^2\omega/(\omega^2-U^2/4)\).  The expression holds
at \emph{any} \(\gamma/U\) and any integer \(\bar n\), and
\(\Sigma_{c_1,i}\) splits into the configuration-independent mean
\(\bar\Sigma\) and the fluctuation \(\delta\Sigma_i\) proportional to
\(S_i|_s\).  In a spin-rotation-symmetric phase the fluctuation
averages to zero over the spin configurations,
\(\langle S_i\rangle=0\).  For \(\nu\neq0\) both statements
presuppose the filling-centred charging term, i.e.\ \(\kappa=1\) in
Eq.~\eqref{eq:HUkappa}.

Setting \(S_i\to0\) in the metric recovers the canonical
effective theory of the previous subsection.  That theory therefore
reproduces the \emph{mean} \(\bar\Sigma(\omega)\) exactly.  But it
carries no coupling to \(S_i\), so it misses the fluctuation
\(\delta\Sigma_i\) entirely.  The missing piece defines an effective
Kondo coupling between \(c_1\) and the local moments.  Because
\(\delta\Sigma_i\) is frequency dependent, the coupling is naturally
written at the level of the action:
\begin{equation}
\delta S_K=\sum_i\int\frac{\dd\omega}{2\pi}\,
\bar c_{1i}(\omega)\,J_K^c(\omega)\,S_i\,c_{1i}(\omega),
\qquad
J_K^c(\omega)=\frac{\gamma^2U}{U^2/4-\omega^2} ,
\label{eq:JKomega}
\end{equation}
where \(\bar c_1 S_i c_1=\sum_{\alpha\beta}\bar c_{1\alpha}
S_i^{\alpha\beta}c_{1\beta}\) and \(S_i\) is promoted from the frozen
matrix \(S_i|_s\) back to the spin operator of
Eq.~\eqref{eq:Xsplit}---an identification exact at a single insertion,
since Eq.~\eqref{eq:configSigmaA} is linear in \(S_i|_s\)
configuration by configuration.  A single insertion of
Eq.~\eqref{eq:JKomega} reproduces \(\delta\Sigma_i(\omega)\)
identically, at every frequency and every \(\gamma/U\).

Two limits then produce the effective Kondo model of the main text.  First, for
\(|\omega|\ll U/2\), we drop the $\omega$ dependence for simplicity and use
\(J_K^c(\omega)=J_K^c(0)\bigl[1+O(4\omega^2/U^2)\bigr]\) with
\begin{equation}
J_K^c(0)=\frac{4\gamma^2}{U}\equiv\kappa(0).
\label{eq:kappazerodef}
\end{equation}
Second, projecting onto
the low-energy band \(\chi\) of the charge sector attaches the
\(c_1\) amplitude of that band to each leg,
\(c_{1\bm k\alpha}\to-\alpha F(k)\,\chi_{\bm k\alpha}\).   This gives the
momentum-space Kondo model
\begin{equation}
H_K=J_K\sum_{\bm k\bm k'}\sum_{\alpha\beta}
F(k)\,F(k')\,
\chi^{\dagger}_{\bm k\alpha}\,S^{\alpha\beta}_{\bm k-\bm k'}\,
\chi_{\bm k'\beta},
\qquad
J_K=\kappa(0)\,\alpha^2=\frac{\gamma^2U}{\gamma^2+U^2/4},
\label{eq:HKapp}
\end{equation}
where \(S^{\alpha\beta}_{\bm q}=N_M^{-1}\sum_ie^{-\ii\bm q\cdot\bm
R_i}S_i^{\alpha\beta}\) is the moment at transferred momentum \(\bm
q\), \(\alpha=(U/2)/\sqrt{\gamma^2+U^2/4}\) is the \(c_1\) weight of
\(\chi\), and
\begin{equation}
F(k)=1-\frac{4|\Phi(k)|^2}{U^2}
=\frac{\kstar^2}{\kstar^2+k^2},
\qquad
\kstar=\frac{\sqrt{\gamma^2+U^2/4}}{v_*},
\label{eq:Fkapp}
\end{equation}
is the \(c_1\)-weight form factor of the low-energy band.  This is
Eq.~\eqref{eq:HK} of the main text, with the same form factor.

The spin operator entering Eq.~\eqref{eq:HKapp} can be
represented by Abrikosov fermions \(\psi'_{i\alpha}\): the
\(\mathcal M\)-dimensional multiplet is exactly the fixed-number
subspace of an \(N_f\)-flavor Fock space, so
\begin{equation}
X_i^{\alpha\beta}=\psi'^{\dagger}_{i\alpha}\psi'_{i\beta},
\qquad
S_i^{\alpha\beta}=\psi'^{\dagger}_{i\beta}\psi'_{i\alpha}
-\frac{\bar n}{N_f}\,\delta_{\alpha\beta}\,\identity_{\mathcal M},
\qquad
\sum_\alpha\psi'^{\dagger}_{i\alpha}\psi'_{i\alpha}=\bar n ,
\label{eq:appmoment}
\end{equation}

In this representation the Kondo term is an explicit four-fermion
vertex.  Let us define the filtered orbital
\(\chi^{F}_{i\alpha}=N_M^{-1/2}\sum_{\bm k}F(k)\,
e^{\ii\bm k\cdot\bm R_i}\chi_{\bm k\alpha}\), where $N_M$ is
the number of sites.  It is not canonically normalized:
\(\{\chi^F_{i\alpha},\chi^{F\dagger}_{i\alpha}\}
=N_M^{-1}\sum_{\bm k}F(k)^2<1\).  No step below uses a canonical
anticommutator for \(\chi^F\); it only bookkeeps the vertex, and
all traces are evaluated with the canonical \(\chi_{\bm k}\).
Then Eq.~\eqref{eq:HKapp} is local,
\begin{equation}
H_K=J_K\sum_i\sum_{\alpha\beta}
\chi^{F\dagger}_{i\alpha}
\Bigl(\psi'^{\dagger}_{i\beta}\psi'_{i\alpha}
-\frac{\bar n}{N_f}\,\delta_{\alpha\beta}\Bigr)
\chi^{F}_{i\beta}.
\label{eq:HKlocal}
\end{equation}
Normal-ordering the moment factor,
\(\psi'^{\dagger}_{i\beta}\psi'_{i\alpha}
=\delta_{\alpha\beta}-\psi'_{i\alpha}\psi'^{\dagger}_{i\beta}\),
collapses the double flavor sum onto a single SU(\(N_f\))-singlet
channel:
\begin{equation}
H_K=J_K\sum_i\Bigl[
\Bigl(1-\frac{\bar n}{N_f}\Bigr)\,n^{\chi F}_i
-\tilde{\mathcal Q}^{\dagger}_i\tilde{\mathcal Q}_i\Bigr],
\qquad
\tilde{\mathcal Q}_i=\sum_\alpha
\psi'^{\dagger}_{i\alpha}\chi^{F}_{i\alpha},
\label{eq:HKsinglet}
\end{equation}
with \(n^{\chi F}_i=\sum_\alpha\chi^{F\dagger}_{i\alpha}
\chi^{F}_{i\alpha}\): the coupling is attractive in the singlet
hybridization channel \(\tilde{\mathcal Q}\) for the antiferromagnetic
sign \(J_K>0\)---which is the sign derived here,
Eq.~\eqref{eq:HKapp} giving \(J_K>0\) at every \(\gamma/U\)---and the
residual potential is cancelled identically by the
Hartree contraction of
\(\tilde{\mathcal Q}^{\dagger}\tilde{\mathcal Q}\), since
\(\langle\psi'_{i\alpha}\psi'^{\dagger}_{i\beta}\rangle
=(1-\bar n/N_f)\delta_{\alpha\beta}\) in the unpolarized multiplet.
Appendix~\ref{app:meanfield} builds the Kondo mean field from exactly
this form.  In the projective limit \(\gamma\gg U\) the dressing
simplifies further: \(\alpha\to0\), so the low-energy field is the
orthogonal fermion itself, \(\chi\to\psi\), while
\(J_K\to U\) and \(\kstar\to\gamma/v_*\), leaving
\begin{equation}
H_K\;\xrightarrow[\gamma\gg U]{}\;
U\sum_i\Bigl[\Bigl(1-\frac{\bar n}{N_f}\Bigr)n^{\psi F}_i
-\tilde{\mathcal Q}^{\dagger}_i\tilde{\mathcal Q}_i\Bigr],
\qquad
\tilde{\mathcal Q}_i=\sum_\alpha\psi'^{\dagger}_{i\alpha}
\psi^{F}_{i\alpha}:
\label{eq:HKprojective}
\end{equation}
the Kondo scale is \(U\) itself and the itinerant carriers are dominated by the orthogonal fermion $\psi$.

\section{Slave-rotor formulation: the Gaussian theory without valence
truncation}
\label{app:rotor}

The construction of Appendix~\ref{app:ancilla} restricts the valence
to \(\{\bar n,\bar n\pm1\}\) and represents the two allowed
transitions by the fermions \(\tilde h,d\).  This appendix
rebuilds the same theory from a slave-rotor representation in which
no truncation is needed: all valence sectors are kept, the doublon
and holon operators arise as \emph{links} of the rotor, and the links
organize into a tower of doublon--holon pairs \((d_l,h_l)\),
\(l=0,1,2,\ldots\).  The orthogonal fermion
\(\psi\) acquires small weights also for the \(l\ge1\) pairs at
\(\pm U(l+\tfrac12)\) which were ignored in our previous theory keeping only the $l=0$ pair.

\subsection{The representation}

Write the \(f\) electron as a rotor times a spinon,
\begin{equation}
f^{\dagger}_{i\alpha}=e^{\ii\theta_i}\,\psi'^{\dagger}_{i\alpha},
\qquad
H_U=\frac U2\sum_i\hat L_i^2,
\qquad
\hat L_i=\sum_\alpha\psi'^{\dagger}_{i\alpha}\psi'_{i\alpha}-\bar n,
\label{eq:rotorrep}
\end{equation}
with \([\theta_i,\hat L_j]=\ii\delta_{ij}\) and
\(e^{\ii\theta}|\ell\rangle=|\ell+1\rangle\) on the rotor ladder
\(\ell\in\mathbb Z\).  The rotor coordinate \(\hat L\) carries
\emph{all} of the site's charge: \(\hat L=n^f-\bar n\), while the spinons
\(\psi'\) carry the spin. Restricted to the multiplet sector
\(\ell=0\), bilinears of $\psi' $ are the spin operators in
Eq.~\eqref{eq:appmoment}.  The representation is exact with the constraint $n_{i;\psi'}= n_{i;f}= \hat L_i+\bar n$.

\subsection{Link operators and the exact algebra}

Decompose the rotor raising operator into its links,
\(e^{\ii\theta}=\sum_\ell|\ell+1\rangle\langle\ell|\), and define one
composite fermion per link and flavor:
\begin{equation}
D^{\dagger}_{\ell,i\alpha}
\equiv|\ell+1\rangle\langle\ell|_i\otimes\psi'^{\dagger}_{i\alpha},
\qquad
f^{\dagger}_{i\alpha}=\sum_\ell D^{\dagger}_{\ell,i\alpha}.
\label{eq:rotorlinks}
\end{equation}
The \(\ell=0\) link is the doublon operator of the main text
(\(\bar n\to\bar n+1\)), the \(\ell=-1\) link is the holon transition
(\(\bar n-1\to\bar n\), i.e.\ \(\tilde h^{\dagger}\)), and
\(\ell=1,-2,\ldots\) are the defect-sector transitions removed by
\(P_G\).  The two-sided index \(\ell\in\mathbb Z\) thus folds at the
reference into one doublon-type link (\(\ell=l\ge0\)) and one
holon-type link (\(\ell=-l-1\)) at every rung distance \(l\); this
pair structure is made explicit below.  Explicitly, for
negative index the link is a generalized holon: writing
\(\ell=-l-1\) with \(l\ge0\),
\(D_{-l-1,i\alpha}=P_{i;\bar n-l-1}\,f_{i\alpha}\,P_{i;\bar n-l}
\equiv H^{(l)\dagger}_{i\alpha}\) creates the removal quantum
\(\bar n-l\to\bar n-l-1\), and \(H^{(0)}=H\) is the holon of
Eq.~\eqref{eq:Hubbardtransitions}.  Because rotor and spinons
commute, the algebra follows in one line:
\begin{equation}
\bigl\{D_{\ell,i\alpha},D^{\dagger}_{\ell',i\beta}\bigr\}
=\delta_{\ell\ell'}\Bigl[P_{\ell}\,\delta_{\alpha\beta}
+\bigl(P_{\ell+1}-P_{\ell}\bigr)\,
\psi'^{\dagger}_{i\beta}\psi'_{i\alpha}\Bigr],
\label{eq:rotoralgebra}
\end{equation}
with \(P_\ell=|\ell\rangle\langle\ell|\).
Separated into the two sides of the ladder, with
\(H^{(l)\dagger}=D_{-l-1}\) as above, Eq.~\eqref{eq:rotoralgebra}
reads
\begin{align}
\{D_{l,i\alpha},D^{\dagger}_{l,i\beta}\}
&=P_{l}\,\delta_{\alpha\beta}
+\bigl(P_{l+1}-P_{l}\bigr)\,\psi'^{\dagger}_{i\beta}\psi'_{i\alpha},
\notag\\
\{H^{(l)}_{i\alpha},H^{(l)\dagger}_{i\beta}\}
&=P_{-l-1}\,\delta_{\alpha\beta}
+\bigl(P_{-l}-P_{-l-1}\bigr)\,\psi'^{\dagger}_{i\alpha}\psi'_{i\beta},
\notag\\
\{D_{l,i\alpha},H^{(l')}_{i\beta}\}&=0 ,
\label{eq:rotoralgebraDH}
\end{align}
for \(l,l'\ge0\).  The flavor bilinear is transposed between the
first two lines.  This is the particle--hole transformation at work:
\(H^{(l)\dagger}\) is a link \emph{annihilation} operator.  The
third line is exact because the two sides live on different links.
On the links \(l=0\) this \emph{derives} the deformed algebra of
Appendix~\ref{app:ancilla}.  Different links anticommute, as the
\(\delta_{\ell\ell'}\) of Eq.~\eqref{eq:rotoralgebra} shows.

\subsection{Canonical link fermions and the untruncated Gaussian
theory}

Summing
Eq.~\eqref{eq:rotoralgebra} over the links telescopes,
\(\{f_{i\alpha},f^{\dagger}_{i\beta}\}
=\delta_{\alpha\beta}\sum_\ell P_\ell=\delta_{\alpha\beta}\): the
electron is the \emph{equal-amplitude} sum of all links and is
exactly canonical.  The untruncated Hamiltonian therefore is in the form
\begin{equation}
H=\sum_i\frac{U}{2}\hat L_i^{2}
+\gamma\sum_{i\alpha}\Bigl[c^{\dagger}_{1,i\alpha}
\sum_\ell D_{\ell,i\alpha}+{\rm H.c.}\Bigr]+H_c ,
\label{eq:rotorexact}
\end{equation}
with \(H_c\) the conduction bands from $c_1,c_2$.   The diagonal of Eq.~\eqref{eq:rotoralgebra}
is
\begin{equation}
\hat w_{\ell,i\alpha}\equiv
\bigl\{D_{\ell,i\alpha},D^{\dagger}_{\ell,i\alpha}\bigr\}
=P_{\ell}\,\bigl(1-\hat n'_{i\alpha}\bigr)
+P_{\ell+1}\,\hat n'_{i\alpha} ,
\label{eq:rotorwop}
\end{equation}
  It is an orthogonal
projector, \(\hat w^2=\hat w\), with \(D_{\ell\alpha}=\hat
w_{\ell\alpha}D_{\ell\alpha}=D_{\ell\alpha}\hat w_{\ell\alpha}\). We also have
\(\sum_\ell\hat w_{\ell,i\alpha}=1\) as an operator identity.

Similar to the previous approach, we can obtain canonical fermion on average if each site is in a mixed state ensemble with vanishing magnetic order.  Let
\(p_\ell=\langle P_\ell\rangle\) be the probability that the site is
in valence sector \(\ell\)  and \(x=\bar n/N_f\). In the free moment ensemble,
In sector \(\ell\) the conditional occupation is
\(\langle\hat n_\alpha\rangle=(\bar n+\ell)/N_f\), so
\begin{equation}
w_\ell=\bigl\langle\hat w_{\ell,i\alpha}\bigr\rangle
=p_\ell\Bigl(1-\frac{\bar n+\ell}{N_f}\Bigr)
+p_{\ell+1}\,\frac{\bar n+\ell+1}{N_f} ,
\label{eq:rotormetric}
\end{equation}
and \(\tilde d_{\ell\alpha}=D_{\ell\alpha}/\sqrt{w_\ell}\)
are canonical on average.  We use this exact metric throughout.
Replacing the conditional occupations by the unconditional mean
\(x\) gives the simpler
\begin{equation}
w_\ell\simeq p_\ell(1-x)+p_{\ell+1}\,x ,
\label{eq:rotormetricfixedx}
\end{equation}
which coincides with Eq.~\eqref{eq:rotormetric} at \(\ell=0\)
up to the \(p_1/N_f\) shift of the second term, and errs at
relative order \((\ell+1)/N_f\) on the ladder.  On the
defect-free reference of the main text (\(p_0=1\)) the two forms
coincide, so Eq.~\eqref{eq:H0} is untouched.  The energy of a
link quantum is the atomic ladder step,
\(\varepsilon_\ell=E_{\ell+1}-E_\ell=U(\ell+\tfrac12)\).

The doublon--holon pairs are now just the normalized \(D\)
and \(H\) links.  For each rung distance \(l=0,1,2,\ldots\),
\begin{equation}
d_{l,i\alpha}\equiv\frac{D_{l,i\alpha}}{\sqrt{w_{\ell=l}}},
\qquad
\tilde h_{l,i\alpha}\equiv h^{\dagger}_{l,i\alpha}
\equiv\frac{H^{(l)\dagger}_{i\alpha}}{\sqrt{w_{\ell=-l-1}}},
\qquad
\varepsilon_l=U\bigl(l+\tfrac12\bigr).
\label{eq:rotorpairs}
\end{equation}
Here \(d^{\dagger}_{l}\) creates the addition quantum
\(\bar n+l\to\bar n+l+1\), and \(h^{\dagger}_{l}\) creates the
removal quantum \(\bar n-l\to\bar n-l-1\).  Both cost the same
atomic step \(\varepsilon_l>0\).  The \(l=0\) pair is the \((d,\tilde h)\) of the
main text; the \(l\ge1\) pairs are the defect-sector channels removed
by \(P_G\).  In pair form the weights Eq.~\eqref{eq:rotormetric} read
\begin{equation}
w_{d_l}=p_{l}\Bigl(1-x-\frac{l}{N_f}\Bigr)
+p_{l+1}\Bigl(x+\frac{l+1}{N_f}\Bigr),
\qquad
w_{h_l}=p_{-l-1}\Bigl(1-x+\frac{l+1}{N_f}\Bigr)
+p_{-l}\Bigl(x-\frac{l}{N_f}\Bigr),
\label{eq:rotorpairweights}
\end{equation}
equal at neutrality,
\(w_{d_l}=w_{h_l}
=p_l(\tfrac12-\tfrac{l}{N_f})
+p_{l+1}(\tfrac12+\tfrac{l+1}{N_f})\), by
particle--hole symmetry.  Promoting every link to an elementary
canonical fermion---the register liberation of
Appendix~\ref{app:ancilla}, now applied to all links---gives the
untruncated Gaussian theory, which is literally a tower of copies of
Eq.~\eqref{eq:H0}, one \((d_l,h_l)\) pair per rung distance:
\begin{equation}
H_G=\sum_{i,\alpha}\sum_{l\ge0}
\varepsilon_l\bigl(n_{d_l,i\alpha}+n_{h_l,i\alpha}\bigr)
+\gamma\sum_{i\alpha}\Bigl[c^{\dagger}_{1,i\alpha}
\sum_{l\ge0}\bigl(\sqrt{w_{d_l}}\;d_{l,i\alpha}
+\sqrt{w_{h_l}}\;\tilde h_{l,i\alpha}\bigr)
+{\rm H.c.}\Bigr],
\label{eq:rotorHG}
\end{equation}
with \(n_{h_l}=1-n_{\tilde h_l}\): every quantum, doublon- or
holon-type, costs \(+\varepsilon_l>0\) relative to the defect-free
reference (all \(d_l\) empty, all \(\tilde h_l\) filled), and the
equal-amplitude electron becomes
\(f_{i\alpha}=\sum_{l\ge0}(\sqrt{w_{d_l}}\,d_{l,i\alpha}
+\sqrt{w_{h_l}}\,\tilde h_{l,i\alpha})\).  Keeping only the \(l=0\)
pair with \(p_0\simeq1\) reproduces \(w_{d_0}=1-x\), \(w_{h_0}=x\)
and hence Eq.~\eqref{eq:H0} exactly.
The \(l=1\) pair carries weights
\(w_{d_1}\simeq p_1(1-x-\tfrac1{N_f})\) and
\(w_{h_1}\simeq p_{-1}(1-x-\tfrac1{N_f})\): the defect-sector
valence probabilities, reduced by their conditional occupations.  The \(\pm3U/2\) channels thus enter as dynamical
fields whose weights are exactly the small parameter of the defect
corrections.  In practice, the weight $w_l$ and $p_l$ should be decided self consistently.   \(H_G\)
determines the valence distribution \(p_\ell\), the distribution
determines the weights \(w_\ell\), and the weights then enter
\(H_G\)  The weights are therefore not
adjustable couplings but ground-state expectation values of the
projector Eq.~\eqref{eq:rotorwop}, decided dynamically.  They obey
the sum rule
\begin{equation}
\sum_{l\ge0}\bigl(w_{d_l}+w_{h_l}\bigr)
=\sum_\ell\Bigl[p_\ell\Bigl(1-\frac{\bar n+\ell}{N_f}\Bigr)
+p_{\ell+1}\,\frac{\bar n+\ell+1}{N_f}\Bigr]=1 ,
\label{eq:rotorsumrule}
\end{equation}
which follows from the operator identity
\(\sum_\ell\hat w_{\ell,i\alpha}=1\).

\begin{figure}[t]
\includegraphics[width=0.85\textwidth]{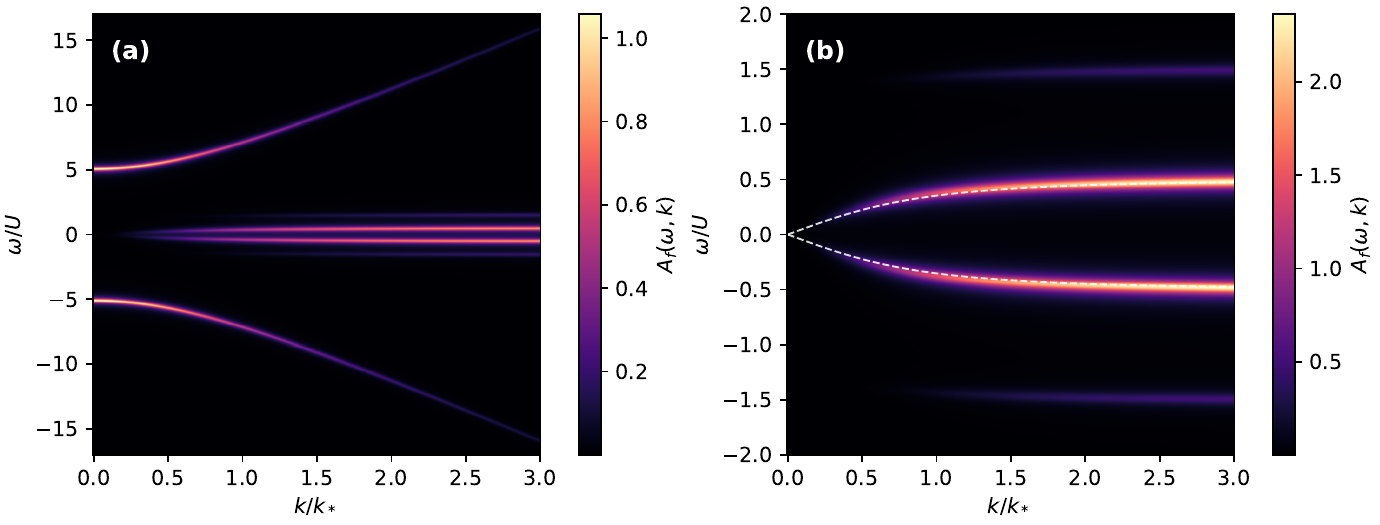}
\caption{\(f\) spectral function \(A_f(\omega,\bm k)\) of the
untruncated Gaussian theory Eq.~\eqref{eq:rotorHG} at
\(\gamma=5U\), \(M=0\), pairs \(l\in\{0,1\}\).  (a) Full window: the bright pair at
\(\pm\sqrt{\gamma^2+v_*^2k^2}\) carries most of the \(f\)
weight.  (b) Low-energy zoom: the Hubbard branch
\(\pm|\Phi(k)|\to\pm U/2\) (white dashed: truncated theory) and
the weak satellite pair at \(\pm3U/2\).}
\label{fig:rotorspec}
\end{figure}

Figure~\ref{fig:rotorspec} shows the \(f\) spectral function
\(A_f(\omega,\bm k)\) at \(\gamma=5U\) at the twist angle \(\theta=1.05^\circ\), \(M=0\) for the charge neutrality, including 
\(l\in\{0,1\}\). From the self-consistent calculation with the
exact metric Eq.~\eqref{eq:rotormetric} we find weights
(\(w_{d_0}=w_{h_0}=0.401\), \(w_{d_1}=w_{h_1}=0.098\)), with
valence distribution \((p_0,p_1,p_2)=(0.556,0.197,0.032)\).  One can see there is a pair of bands at energy $E \approx \pm \frac{3U}{2}$ with small weights.

\subsection{The orthogonal fermion and the dark zero mode in
rotor language}

The physical electron is one specific combination of link fermions,
\(f_{i\alpha}=\sum_{l}(\sqrt{w_{d_l}}\,d_{l,i\alpha}
+\sqrt{w_{h_l}}\,\tilde h_{l,i\alpha})\); the orthogonal fermions are
the \emph{metric-orthogonal complement} of this combination in link
space.  With the dominant \(l=0\) pair alone,
\begin{equation}
f_{0,i\alpha}=\sqrt{w_{h_0}}\,\tilde h_{0,i\alpha}
+\sqrt{w_{d_0}}\,d_{0,i\alpha},
\qquad
\psi_{i\alpha}=\sqrt{w_{h_0}}\,d_{0,i\alpha}
-\sqrt{w_{d_0}}\,\tilde h_{0,i\alpha},
\label{eq:rotorpsi}
\end{equation}
which reproduces Eq.~\eqref{eq:PGfgeneral} with
\(w_{h_0}=x\), \(w_{d_0}=1-x\), and at \(\nu=0\)
(\(w_{d_0}=w_{h_0}=\tfrac12\)) is
\(f_0=(\tilde h+d)/\sqrt2\),
\(\psi=(d-\tilde h)/\sqrt2\) of the main text.  \(\psi\) is 
the sign-alternating combination of links---compactly,
\(\psi\propto e^{-\ii\theta}\,{\rm sgn}(\hat L-\tfrac12)\,\psi'\) in
the truncated space.  With more pairs kept there are more dark
combinations.  The link space holds \(2n_{\rm pairs}\) orbitals per
flavor.  The hybridization in Eq.~\eqref{eq:rotorHG} contains
exactly one combination of them, the electron \(f\) itself.  A link
mode \(b_v=\sum_\ell v_\ell\tilde d_\ell\) therefore decouples
iff its anticommutator with the electron vanishes,
\(\{b_v,f^{\dagger}\}=\sum_\ell\sqrt{w_\ell}\,v_\ell=0\).
This is one linear condition on \(2n_{\rm pairs}\) amplitudes, so
the dark space has dimension \(2n_{\rm pairs}-1\).

At \(\bm k=0\), there is one zero-energy dark mode.  In
the projective limit \(\gamma\gg U\) its wavefunction is
\begin{equation}
\psi_{0,\alpha}(\bm k{=}0)
=\mathcal N\sum_{l\ge0}\frac{1}{l+\tfrac12}\,
\bigl[\sqrt{w_{d_l}}\,d_{l,\alpha}(\bm 0)
-\sqrt{w_{h_l}}\,\tilde h_{l,\alpha}(\bm 0)\bigr],
\qquad
\mathcal N=\Bigl[\sum_{l\ge0}
\frac{w_{d_l}+w_{h_l}}{(l+\tfrac12)^{2}}\Bigr]^{-1/2},
\label{eq:rotorpsi0}
\end{equation}
where \(d_{l,\alpha}(\bm 0)=N_M^{-1/2}\sum_i d_{l,i\alpha}\) is
the momentum-zero link mode, and likewise for
\(\tilde h_{l,\alpha}(\bm 0)\).  At general \(\gamma\) the exact zero mode acquires a \(c_1\)
admixture, \(\sqrt Z\,c_1-\sqrt{1-Z}\,\psi_0\) with
\(Z=[1+\gamma^2\sum_\ell w_\ell/\varepsilon_\ell^{2}]^{-1}\).

In microscopic language the same operator is 
\begin{equation}
\psi_{0,\alpha}(\bm k{=}0)
=\frac{\mathcal N}{\sqrt{N_M}}\sum_i
\bigl(\delta\hat n_i+\tfrac12\bigr)^{-1}
f_{i\alpha}.
\label{eq:rotorpsi0micro}
\end{equation}

It satisfies the identity
\([H_U,\psi_0(\bm k=0)]=-\mathcal NU\,f(\bm k=0)\) and thus vanishes in the projective limit.  In the tower it differs
from the first Krylov vector
\(F=\{f,\delta\hat n\}\propto[H_U,f]\), which weights the pair
\(l\) by \(\varepsilon_l\) instead of \(1/\varepsilon_l\).

\section{Density of doublon and holon excitations}
\label{app:mixedvalence}
\label{app:fcs}

This appendix quantifies the doublon and holon content of the
mixed-valence Mott state.  We take the THFM active band at the
chiral twist angle \(\theta=1.081^\circ\), where \(M=0\), with
\(U=30\)~meV and \(N_f=8\); the hybridization is taken deep in
the projective regime, \(\gamma=3U=90\)~meV, and the patch
fraction is \(s^2=\pi\kstar^2/\Omega_M=0.04\), with
\(\Omega_M\) the Brillouin-zone area.  We diagonalize Eq.~\eqref{eq:Hhd} at
each \(\bm k\) and fill the states below the chemical potential,
which fixes the filling \(\nu\).  The occupations of the two charge channels
are
\begin{equation}
n_d(\bm k)=\bigl\langle d^{\dagger}_{\bm k}
d_{\bm k}\bigr\rangle,
\qquad
n_h(\bm k)=1-\bigl\langle\tilde h^{\dagger}_{\bm k}
\tilde h_{\bm k}\bigr\rangle.
\label{eq:ndnh}
\end{equation}

\begin{figure*}
\includegraphics[width=\textwidth]{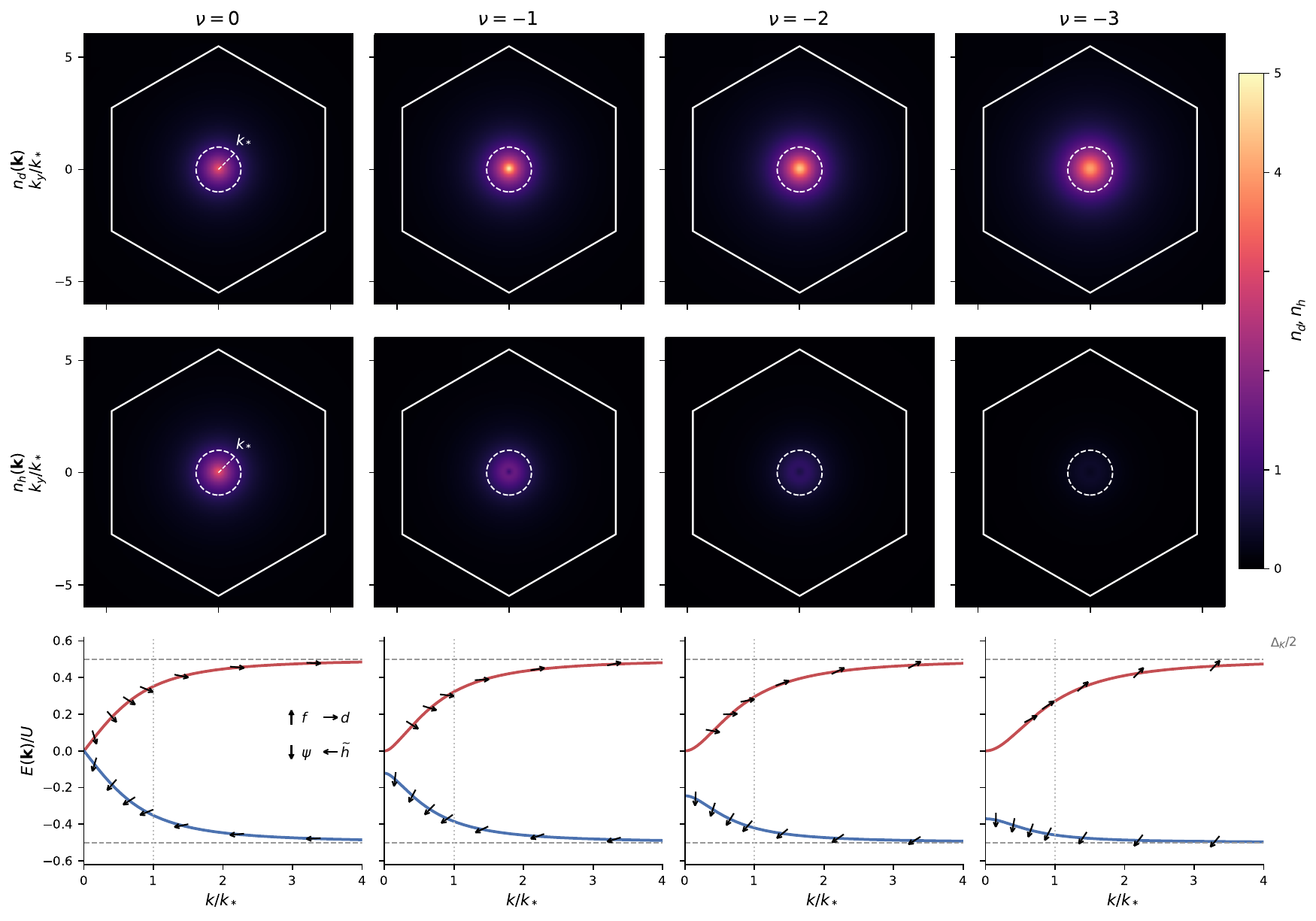}
\caption{Mixed valence of the Mott state from the Gaussian theory
Eq.~\eqref{eq:Hhd}, for \(\nu=0,-1,-2,-3\) (columns) at
\(\gamma/U=3\), \(s^2=0.04\), \(N_f=8\).  Top and middle
rows: momentum maps of the doublon and holon occupations
\(n_d(\bm k)\) and \(n_h(\bm k)\) [Eq.~\eqref{eq:ndnh}],
summed over flavors, on the moir\'e Brillouin zone (white hexagon;
dashed circle \(k=\kstar\); shared colour scale).  Bottom row:
the two low-energy bands, degenerate at \(\Gamma_M\) at
\(\nu=0\), split by \(\delta_\nu\) away from it, and
saturating at \(\pm U/2\) (dashed) beyond the patch.
Arrows: the \((f,\psi)\) content of the band eigenvector,
Eq.~\eqref{eq:arrowangle} (legend in the first panel); no arrow is
drawn where the band is \(c\)-dominated.}
\label{fig:mixedvalence}
\end{figure*}

Figure~\ref{fig:mixedvalence} shows the momentum-resolved
\(n_d(\bm k)\) and \(n_h(\bm k)\), and the two low-energy
bands, for \(\nu=0,-1,-2,-3\).  Doublon and holon excitations are concentrated in a region around $\bm k=0$ and as a result the low energy bands involve a mixture of them: the orthogonal fermion $\psi(\bm k)$. In the region $|\bm k|\gg\kstar$, the ground state has a fixed configuration $n^f_i = \bar n$ and therefore the UHB and the LHB are dominated by $d$ and $\tilde h$ respectively.
The evolution is labeled by the arrows in the bottom row.  The link
content of a band is decomposed in the \((f,\psi)\) basis, with
\(f=\sqrt{w_{d}}\,d+\sqrt{w_{h}}\,\tilde h\) the electron and
\(\psi=\sqrt{w_{h}}\,d-\sqrt{w_{d}}\,\tilde h\) the
orthogonal fermion, and parametrized by an angle:
\begin{equation}
u(\bm k)=\cos\tfrac{\theta_{\bm k}}2\,f_{\bm k}
+\sin\tfrac{\theta_{\bm k}}2\,\psi_{\bm k},
\qquad
\tan\theta_{\bm k}
=\frac{2\,u_f\,u_{\psi}}{u_f^{2}-u_{\psi}^{2}}.
\label{eq:arrowangle}
\end{equation}
The arrow points at angle \(\theta_{\bm k}\) measured from the
vertical.  Up (\(\theta=0\)) is the electron \(f\).  Down
(\(\theta=\pi\)) is the orthogonal fermion \(\psi\).  A legend
is drawn in the first panel.  At \(\nu=0\) the far-zone Hubbard
branches are the horizontal arrows: the pure doublon (right) and the
pure holon (left).  This convention is filling independent, and it
displays the electron content of the band directly:
\begin{equation}
Z_{\bm k}=r_{\bm k}\cos^{2}\frac{\theta_{\bm k}}2
\label{eq:arrowweight}
\end{equation}
at every \(\nu\), with \(r_{\bm k}=u_f^{2}+u_{\psi}^{2}\) the
band's total link fraction.  A down arrow is dark.  An up arrow
carries the full link weight.  A horizontal arrow carries half.
This is the weight transfer seen in Fig.~\ref{fig:rotorspec}, and
away from neutrality the Hubbard-branch arrows tilt toward \(f\) on
the doublon side, \(Z\to r_{\bm k}w_{d}\), and away from it on
the holon side, \(Z\to r_{\bm k}w_{h}\).  At the patch boundary
\(k=\kstar\) the values are: \(Z_{\kstar}=0.25\) for both bands
at \(\nu=0\), and \((0.32,0.18)\), \((0.38,0.12)\),
\((0.44,0.06)\) for the (upper, lower) band at \(\nu=-1,-2,-3\).
The sum stays \(\simeq\tfrac12\) throughout, while doping shifts
the weight to the doublon-dominated upper branch.

\section{Scattering rate from $O(J_K^2)$ order}
\label{app:born}

This appendix derives the self-energy of the low-energy field
\(\chi\) (the orthogonal fermion dressed by the projection,
\(\chi\to\psi\) in the projective limit) at order \(J_K^2\),
from scattering off free, unpolarized, quasi-static moments: the
regime \(T\gg T_K\).  We work at charge neutrality,
\(\bar n=N_f/2\), keep \(N_f\) general so that the flavor count
appears as an explicit prefactor, and express every rate through the
patch fraction \(s^2=\pi\kstar^2/\Omega_M\).  We also introduce a single
dimensionless factor
\begin{equation}
\zeta\equiv\frac{J_K}{U}=\frac{\gamma^2}{\gamma^2+U^2/4}
\;\xrightarrow[\gamma\gg U]{}\;1.
\label{eq:zetafactor}
\end{equation}
At the end of the
appendix we will compare with the recent results for the scattering rate based on the Hubbard-III approximation or controlled diagram calculations without a Kondo interpretation
\cite{VituriBerg2026,Wei2026,Hu2026,Nosov2026}.

 For independent
unpolarized SU(\(N_f\)) moments at \(\bar n=N_f/2\), the
contraction over final moment flavors is
\begin{equation}
\sum_r
\left\langle
S_i^{sr}S_j^{rs'}
\right\rangle_T
=C_{N_f}\,\delta_{ij}\delta_{ss'},
\qquad C_{N_f}=\frac{N_f+1}{4},
\label{eq:THFMspinvariance}
\end{equation}
so that \(C_2=3/4\) recovers the familiar spin-\(\tfrac12\)
value, and \(C_8=9/4\) for the TBG multiplet.  With the lattice
normalization \(N_M^{-1}\sum_{\bm q}=A_M\int\dd^2q/(2\pi)^2\)
(\(A_M\Omega_M=(2\pi)^2\), with \(A_M\) the moir\'e cell area)
the Born self-energy for one orbital block is
\begin{equation}
\left[\Sigma_\chi^R(\bm k,\omega)\right]_{ab}
=C_{N_f}J_K^2\,F(k)^2A_M
\int\frac{\dd^2q}{(2\pi)^2}\,F(q)^2
\left[G_{\chi\chi}^{R,0}(\bm q,\omega)\right]_{ab}.
\label{eq:THFMBorn}
\end{equation}
Here \(J_K=\zeta U\) is the static Kondo coupling of the projected
model, Eq.~\eqref{eq:HKapp}.  
Equation~\eqref{eq:THFMBorn} is the second-order self-energy.

The propagator in Eq.~\eqref{eq:THFMBorn} carries the two
\(f\)-orbital indices \(a,b\), and after the momentum integral it
is diagonal, \(\propto\delta_{ab}\).  At \(M=0\) it is diagonal
from the start:
\begin{equation}
\bigl[G_{\chi\chi}^{R,0}(\bm q,\omega)\bigr]_{ab}
=\delta_{ab}\,
\frac{\omega+\ii0^+}{(\omega+\ii0^+)^2-\Phi_\Gamma^2q^2}.
\label{eq:THFMDiracG}
\end{equation}
At finite \(M\), we have:
\begin{equation}
\int\frac{\dd\theta_{\bm q}}{2\pi}\,
\bigl[G_{\chi\chi}^{R,0}(\bm q,z)\bigr]_{ab}
=\delta_{ab}\,\frac{a_q(z)}{a_q^{2}(z)-b^{2}(z)\,q^{4}},
\qquad
a_q(z)=z\Bigl(1+\frac{\Phi_\Gamma^2q^2}{M^2-z^2}\Bigr),
\quad
b(z)=\frac{\Phi_\Gamma^2M}{M^2-z^2}.
\label{eq:THFMGangular}
\end{equation}
The densities of states quoted below are the imaginary parts of
these expressions integrated over \(q\).
Inside the patch \(F\simeq1\), and the imaginary parts below are
cutoff free; the form factors only regularize the logarithmic real
parts, cutting them at \(\Lambda\lesssim\kstar\).  We therefore
set \(F\to1\) and keep an explicit cutoff \(\Lambda\) in the
logarithms.  Equation~\eqref{eq:THFMBorn} is the starting point for
the two cases below.

At \(M=0\) the patch hosts the chiral Dirac cone.  The \(\chi\)
density of states vanishes linearly,
\(\rho_\chi^{\rm D}(\omega)=|\omega|/(4\pi\Phi_\Gamma^2)\),
so the rate vanishes at the node:
\begin{equation}
\boxed{\;
\Gamma_\chi^{\rm D}(\omega)
=(N_f+1)\,\pi s^2 \zeta^2\,|\omega| .
\;}
\label{eq:THFMDiracRate}
\end{equation}

At finite \(M\) and \(|\omega|\ll|M|\), we have a quadratic band touching for $\chi$:
\begin{equation}
\mathcal A=\frac{\Phi_\Gamma^2}{|M|},
\qquad
\varepsilon_{\rm QBT}(k)=\mathcal A k^2 .
\label{eq:THFMA}
\end{equation}
The local \(\chi\) density of states is now constant,
\(\rho_\chi^{\rm QBT}(0)=|M|/(8\pi\Phi_\Gamma^2)\), so the
momentum integral of Eq.~\eqref{eq:THFMBorn} is finite at zero
energy.  The Kondo model therefore gives
\begin{equation}
\boxed{
\Gamma_\chi^{\rm QBT}(0)
=\frac{C_{N_f}J_K^2A_M}{8\mathcal A}
=\frac{C_{N_f}J_K^2A_M|M|}{8\Phi_\Gamma^2}
=\frac{N_f+1}{2}\,\pi s^2\zeta^2\,|M| .
}
\label{eq:THFMQBTGamma}
\end{equation}
The complete leading low-energy self-energy is
\begin{equation}
\boxed{
\Sigma_\chi^{R,\rm QBT}(z)
=-(N_f+1)\,s^2\zeta^2\,
z\ln\frac{\mathcal A\Lambda^2}{|z|}
-\ii\,\frac{N_f+1}{2}\,\pi s^2\zeta^2\,|M|
+O(z).
}
\label{eq:THFMQBTComplexSigma}
\end{equation}

\emph{Comparison with recent works.}  Our results agree with the
recent works, which rely on the Hubbard-III approximation
\cite{Wei2026} or on controlled diagrammatic expansions
\cite{VituriBerg2026,Hu2026,Nosov2026}, but now we have a clear
physical interpretation in terms of a Kondo coupling.  In
the projective limit \(\zeta\to1\) they reduce to the projected
results of Ref.~\cite{Wei2026}, including the \(N_f+1\)
prefactor.  The \(\gamma^4/(\gamma^2+U^2/4)\) dependence at general
\(\gamma/U\) agrees with both
Refs.~\cite{VituriBerg2026,Hu2026}: the loop expansion of
Ref.~\cite{VituriBerg2026} obtains it with prefactor \(N_f\), the
leading term of its loop counting, and Eq.~(17) of
Ref.~\cite{Hu2026} equals Eq.~\eqref{eq:THFMQBTGamma} exactly.  The result of
Ref.~\cite{Nosov2026} is the \(N_f=2\), \(\zeta=1\) case. To the best of our knowledge, none of these works explicitly interprets the scattering as arising from a Kondo coupling. In our interpretation, the results can be obtained from a simple calculation. And more importantly, with the Kondo interpretation, we know the perturbative calculations would definitely break down at lower temperature and the major focus of this paper is on the Kondo screened phase which is likely beyond the framework of these recent works.

\section{Kondo mean field of the projected model and the screening
scale \texorpdfstring{\(T_K(\theta)\)}{TK(theta)}}
\label{app:meanfield}

So far the moments were treated as free and quasi-static, the
regime \(T\gg T_K\).  Here we take the complementary limit and
carry the emergent Kondo lattice to its saddle point, which fixes
\(T_K\) itself.  The calculation keeps only the three fields of
the projective limit: the active electron \(c\), the orthogonal
fermion \(\chi\) (written \(\psi\) throughout), and the moment
fermion \(\psi'\).  It is formulated in the ancilla framework, in
a unified language for all integer fillings \(\nu=0,-1,-2,-3\).
Its inputs are the four numbers \(M,\gamma,v_*,U\) at each twist
angle; its output is the \(T_K(\theta,\nu)\) of
Fig.~\ref{fig:overview}(b) of the main text.

\subsection{The projected model}
\label{app:mfmodel}

Per valley--spin flavor (four of them, each carrying a two-component
orbital spinor, so \(N_f=8\) in total),
\begin{align}
H={}&\sum_{\bm k}\Big[
c^{\dagger}_{\bm k}\bigl[h_a(\bm k)-\mu_c\bigr]c_{\bm k}
+\Bigl(\frac{\nu U}{8}-\mu_\psi\Bigr)
\psi^{\dagger}_{\bm k}\psi_{\bm k}
+\bigl(c^{\dagger}_{\bm k}\Phi_{\bm k}\psi_{\bm k}
+{\rm H.c.}\bigr)
-\mu_{\psi'}\,\psi'^{\dagger}_{\bm k}\psi'_{\bm k}\Big]
\notag\\
&+J_K\sum_{\bm k\bm k'}F(k)F(k')
\sum_{\alpha\beta}
\psi^{\dagger}_{\bm k\alpha}\,
S^{\alpha\beta}_{\bm k-\bm k'}\,
\psi_{\bm k'\beta} .
\label{eq:mfmodel}
\end{align}
Here \(\Phi_{\bm k}\) is the saturating hybridization of the main
text, \(|\Phi(k)|=(U/2)\,k/\sqrt{k^2+\kstar^2}\), and
\(F(k)=\kstar^2/(\kstar^2+k^2)\) is its form factor
in the main text.  The active electron is not the bare
\(\Gamma_1\oplus\Gamma_2\) doublet away from \(\Gamma_M\): its
orbital content rotates to \(f\)-like on the scale \(\kstar\).
The one-body terms are therefore projected on the band,
\begin{equation}
h_a(\bm k)=m(k)\,\sigma_x,
\quad
m(k)=\sqrt{F^2M^2+[1-F]^2\varepsilon_f^2},
\quad
\varepsilon_f(k)=6t_0\sqrt{J_2^2+J_4^2}\,(ka_M),
\label{eq:activedisp}
\end{equation}
the mass with the \(\Gamma_1\oplus\Gamma_2\) weight \(F(k)\) and
the \(f\)-electron nearest-neighbor hopping \(t_0\) of
Ref.~\cite{Calugaru2023} with the complementary weight
(\(M\) denotes \(|M|\) here and below).  The hopping is
inter-orbital: the particle--hole symmetry of the THF model forbids
an orbital-diagonal \(f\) dispersion, so \(t_0\) enters the mass
channel.  Its form factor
\(t(\bm k)=\sum_{\bm\delta}e^{i\bm k\cdot\bm\delta
-4i\phi_\delta}\) winds relative to the constant mass, so on the
angle average the two terms add in quadrature: \(\varepsilon_f\)
is the root-mean-square of the \(m=2\) and \(m=4\) Bessel
harmonics of \(t(\bm k)\), and it vanishes at \(\Gamma_M\),
where the splitting is already \(M\).  On high-symmetry paths we
use \(|F M+(1-F)t_0t(\bm k)|\) directly; \(t(\bm k)\) vanishes
at \(K_M\).  At \(\nu=0\) the model is exactly particle--hole
symmetric.  We
checked the mass projection against the four-orbital star model of
Appendix~\ref{app:exactsigma}: the \(M\)-induced splitting of the
low-energy bands follows \(MF(k)\) closely at every \(k\), while
an unprojected \(M\) overestimates it by a factor of two and more
beyond \(\kstar\).  Near the magic angle \(t_0\lesssim0.1\)~meV
and \(\varepsilon_f\) is negligible; it grows to \(2.8\)~meV at
\(\theta=1.30^\circ\).  The flavor indices
\(\alpha,\beta\) run over all \(N_f\) components, and
\(S^{\alpha\beta}_{\bm q}
=N_M^{-1}\sum_ie^{-i\bm q\cdot\bm R_i}S_i^{\alpha\beta}\)
is the momentum transform of the moment operator
[Eq.~\eqref{eq:appmoment}].  The Kondo coupling is the static
vertex of Appendix~\ref{app:exactsigma}, projected on the low-energy
band [Eq.~\eqref{eq:HKapp}]:
\begin{equation}
J_K=\zeta U=\frac{\gamma^2U}{\gamma^2+U^2/4},
\label{eq:JKmf}
\end{equation}
with the \(\zeta\) of Eq.~\eqref{eq:zetafactor}; \(J_K=0.79U\)
at the magic angle (\(\theta=1.081^\circ\), where \(M=0\)).
It is filling independent.  The diagonal
\(\nu U/8=(\bar n/N_f-\tfrac12)U\) is the atomic offset of the
orthogonal-fermion level at reference filling \(\bar n=4+\nu\)
[Eq.~\eqref{eq:Ubardelta}].  The derived scales are
\(\kstar=\sqrt{\gamma^2+U^2/4}/v_*\),
\(\alpha=(U/2)/\sqrt{\gamma^2+U^2/4}\),
\(\Phi_\Gamma=\alpha v_*\),
\(2\Phi_\Gamma\kstar=U\), and
\(s^2=\pi\kstar^2/\Omega_M\).

The ancilla construction fixes three densities per site:
\begin{equation}
n_c=4+\nu,
\qquad
n_\psi=4-\nu,
\qquad
n_{\psi'}=4+\nu .
\label{eq:mfconstraints}
\end{equation}
The physical active filling is carried by \(c\); the two ancilla
layers satisfy \(n_\psi+n_{\psi'}=N_f\).  Each density has its
own chemical potential in Eq.~\eqref{eq:mfmodel}: \(\mu_c\),
\(\mu_\psi\) and \(\mu_{\psi'}\) are the Lagrange multipliers
of the three constraints.  All three are determined
self-consistently at each temperature, above and below the
transition.  In the normal state (\(b=0\)) the moment layer
decouples and is flat, so its constraint is solved in closed form:
\begin{equation}
\mu_{\psi'}=T\,\ln\frac{\bar x}{1-\bar x},
\qquad
\bar x=\frac{4+\nu}{8},
\label{eq:momentlevel}
\end{equation}
i.e.\ the moment level sits at \(-\mu_{\psi'}\), above the Fermi
level for \(\nu<0\).  Only this \(b=0\) value enters the onset
calculation below.  Once \(b\neq0\) the condensate hybridizes the
moment level with the charge sector and no closed form exists;
\(\mu_{\psi'}\) is then solved together with \(\mu_c\) and
\(\mu_\psi\).  At \(\nu=0\) particle--hole symmetry gives
\(\mu_c=\mu_\psi=\mu_{\psi'}=0\) exactly; at \(\nu\neq0\)
the shifts are solved numerically.

\subsection{The Kondo term is exactly an attraction in the singlet
channel}

Before any approximation, the vertex can be written as a single
bilinear-squared.  Using Eq.~\eqref{eq:appmoment} at reference
filling \(\bar n\),
\begin{equation}
\sum_{\alpha\beta}\psi^{\dagger}_{i\alpha}
S_i^{\alpha\beta}\psi_{i\beta}
=(1-\bar x)\,n^{\psi}_i-\mathcal Q^{\dagger}_i\mathcal Q_i,
\qquad
\mathcal Q_i\equiv\sum_\alpha
\psi'^{\dagger}_{i\alpha}\psi_{i\alpha},
\label{eq:singletidentity}
\end{equation}
an operator identity, exact in the constrained space at every
integer filling.  Two things follow.  The moment coupling is
\emph{attractive} in the SU(\(N_f\)) singlet channel
\(\mathcal Q\) for the physical antiferromagnetic sign
\(J_K>0\), with a strength independent of \(\nu\).  And the
residual potential \((1-\bar x)J_Kn^{\psi}\) is not a physical
shift: it is cancelled identically by the direct (Hartree)
contraction of \(\mathcal Q^{\dagger}\mathcal Q\), consistent
with \(\langle S_i\rangle=0\) for the unpolarized multiplet.  We
therefore drop both and decouple only the singlet channel.

\subsection{Saddle point}

Write \(b_i=J_K\langle\mathcal Q_i\rangle\), uniform and real:
the global U(1) phase of \(\psi'\) is a gauge freedom of the
Abrikosov representation, which we fix by \(b\ge0\).  The free
energy depends on \(|b|\) only, so the two signs are one saddle,
not two.
With the separable vertex the saddle closes on a single condensate:
\begin{equation}
B=\frac1{N_M}\sum_{\bm k}F(k)\,
\bigl\langle\psi'^{\dagger}\psi_{\bm k}\bigr\rangle,
\qquad
b(\bm k)=J_K\,B\,F(k),
\label{eq:mfhyb}
\end{equation}
a momentum-selective Kondo hybridization concentrated on the patch.
The mean-field Hamiltonian per valley--spin flavor is the
\(6\times6\) matrix
\begin{equation}
H_{\rm MF}(\bm k)=
\begin{pmatrix}
h_a(\bm k)-\mu_c&\Phi_{\bm k}&0\\
\Phi^{\dagger}_{\bm k}&\bigl(\tfrac{\nu U}8-\mu_\psi\bigr)
\sigma_0&b(\bm k)\sigma_0\\
0&b(\bm k)\sigma_0&-\mu_{\psi'}\sigma_0
\end{pmatrix}.
\label{eq:HMF}
\end{equation}

\subsection{Onset condition}

At the transition \(b\to0\).  Linearizing,
\(\langle\psi'^{\dagger}\psi_{\bm k}\rangle
=b(\bm k)K(\bm k)\) with
\begin{equation}
K(\bm k)=\sum_n\bigl|u_{\psi n}(\bm k)\bigr|^2\,
\frac{n_F(-\mu_{\psi'})-n_F[\varepsilon_n(\bm k)]}
{\varepsilon_n(\bm k)+\mu_{\psi'}},
\label{eq:Kbubble}
\end{equation}
where \(n\) runs over the four charge-sector branches at the
constrained levels, \(|u_{\psi n}|^2\) is their
orthogonal-fermion weight, and all energies are measured from the
Fermi level.  At \(\nu=0\), \(\mu_{\psi'}=0\) and the bubble is
\(\tanh(\varepsilon/2T)/2\varepsilon\).  Because the vertex is
separable, the self-consistency closes on the single number
\(B\), and the instability condition is
\begin{equation}
\boxed{\;J_K\,\mathcal K_2(T_K)=1\;,}
\qquad
\mathcal K_j(T)\equiv\frac{4}{N_M}\sum_{\bm k}F(k)^j\,
K(\bm k)
\label{eq:TKcondition}
\end{equation}
(the factor four is the valley--spin count; the orbital sum is
inside \(|u_{\psi n}|^2\)).  With \(F\equiv1\) this is the
textbook \(J_K\mathcal K=1\), i.e.\
\(T_K\simeq\Lambda\,e^{-1/(N_fJ_K\rho)}\), with the flavor
count in the exponent.

At \(\nu\neq0\) the charge sector can be gapped at the Fermi
level.  Inside such a window \(\rho_\psi(E_F)=0\) and screening
is impossible at \(T=0\).  The finite-\(T\) mean field can still
condense once \(T\gtrsim\Delta_g\) populates the bands
thermally, so we use the zero-temperature criterion: wherever the
charge gap \(\Delta_g\) exceeds the nominal \(T_K\), the
moments are unscreened and \(T_K\equiv0\).

\end{document}